\UseRawInputEncoding
\documentclass[superscriptaddress, twocolumn, amsmath, amssymb, aps,pra, notitlepage,longbibliography]{revtex4-2} %preview twocolumn
\usepackage{graphicx,graphics,epsfig,subfigure,times,bm,bbm,amssymb,amsmath,amsfonts,amsthm,mathrsfs,MnSymbol,physics}
\usepackage[matrix,frame,arrow]{xypic}
\usepackage[normalem]{ulem}
\usepackage{slashed}
\usepackage{dcolumn}
\usepackage{tabularx}
\usepackage{array}
\usepackage{amsopn}
\usepackage{color}
\usepackage[usenames,dvipsnames,svgnames,table]{xcolor}
\usepackage[english]{babel}
\usepackage{verbatim}
\usepackage{amsmath}
\definecolor{darkblue}{rgb}{0.0,0.0,0.3}
\usepackage[colorlinks=true,
            linkcolor=red,
            urlcolor= darkblue,
            citecolor=blue]{hyperref}

\newcommand{\bea}{\begin{eqnarray}}
\newcommand{\eea}{\end{eqnarray}}

\begin{document}
%\title{Going beyond Landauer: Information-cost relations from inference based on the maximum entropy principle}
\title{Quantum information-cost relations and fluctuations beyond thermal environments: A thermodynamic inference approach}

\author{Yuanyuan Xiao}
\affiliation{Department of Physics, Institute for Quantum Science and Technology, Shanghai Key Laboratory of High Temperature Superconductors, International Center of Quantum and Molecular Structures, Shanghai University, Shanghai, 200444, China}
\author{Jian-Hua Jiang}
\email{jhjiang3@ustc.edu.cn}
\affiliation{School of Biomedical Engineering, Division of Life Sciences and Medicine, University of Science and Technology of China, Hefei 230026, China}
\affiliation{Suzhou Institute for Advanced Research, University of Science and Technology of China, Suzhou, 215123, China}
\affiliation{School of Physical Sciences, University of Science and Technology of China, Hefei, 230026, China}
\author{Junjie Liu}
\email{jj\_liu@shu.edu.cn}
\affiliation{Department of Physics, Institute for Quantum Science and Technology, Shanghai Key Laboratory of High Temperature Superconductors, International Center of Quantum and Molecular Structures, Shanghai University, Shanghai, 200444, China}

\begin{abstract}
The Landauer's principle, a cornerstone of information thermodynamics, provides a fundamental lower bound on the energetic cost of information erasure in terms of the information content change. However, its traditional formulation is largely confined to systems exchanging solely energy with an ideal thermal bath. In this work, we derive general information-cost trade-off relations that go beyond the scope of Landauer's principle by developing a thermodynamic inference approach based on the maximum entropy principle. These relations require only information about the system and are applicable to complex quantum scenarios involving multiple conserved charges and non-thermal environments. Specifically, we present two key results: (i) In scenarios where only the mean values of observables are accessible, we derive an information-content-informed upper bound on the thermodynamic cost which complements an existing generalized Landauer lower bound. (ii) When second-order fluctuations can also be measured, we obtain an information-content-informed lower bound on the change in variances of observables, thereby extending the Landauer's principle to constrain higher-order fluctuation costs. We numerically validate our information-cost trade-off relations using a coupled-qubit system exchanging energy and excitations, a driven qubit implementing an information erasure process, and a driven double quantum dot system that can operate as an inelastic heat engine. Our results underscore the broad utility of maximum-entropy inference in constraining thermodynamic costs for generic finite-time quantum processes, with direct relevance to quantum information processing and quantum thermodynamic applications.
\end{abstract}

\date{\today}
\maketitle

\section{Introduction}
The relentless miniaturization of information processing devices into the quantum realm has brought the fundamental connection between information and thermodynamics to the forefront~\cite{Parrondo.15.NP,Goold.16.JPA}. Landauer's principle (LP)~\cite{Landauer.61.IBM}, being an information-cost relation that posits a minimal energy cost for erasing one classical bit of information, offers an elegant embodiment of this connection. Recognizing that the LP is closely tied to the Clausius inequality for total entropy production~\cite{Esposito.10.NJP,Reeb.14.NJP,Landi.21.RMP}, subsequent refinements~\cite{Sagawa.09.PRL,Esposito.10.NJP,Hilt.11.PRE,Deffner.13.PRX,Lorenzo.15.PRL,Dago.21.PRL,Riechers.21.PRA,Goold.15.PRL,Esposito.11.EPL,Campbell.17.PRA,Reeb.14.NJP,Browne.14.PRL,Bera.17.NC,Miller.20.PRL,Proesmans.20.PRL,Saito.22.PRL,LeeJ.22.PRL,Dago.22.PRL,Timpanaro.20.PRL,Hsieh.25.PRL,Liu.23.PRAa,LiuJ.24.PRR} have
successfully extended its scope beyond information erasure, establishing it as a fundamental constraint applicable to both classical and quantum systems. Recent experimental demonstrations \cite{Peterson.16.PRSA,Yan.18.PRL,Gaudenzi.18.NP,Aimet.25.NP} have further reinforced the validity of the LP.

While foundational, an important caveat of the LP is its requirement that the system’s environment be in thermal equilibrium at a fixed temperature. This assumption permeates much of the subsequent literature on thermodynamic cost bounds~\cite{Sagawa.09.PRL,Esposito.10.NJP,Hilt.11.PRE,Deffner.13.PRX,Lorenzo.15.PRL,Dago.21.PRL,Riechers.21.PRA,Goold.15.PRL,Esposito.11.EPL,Campbell.17.PRA,Reeb.14.NJP,Browne.14.PRL,Bera.17.NC,Miller.20.PRL,Proesmans.20.PRL,Saito.22.PRL,LeeJ.22.PRL,Dago.22.PRL,Timpanaro.20.PRL,Hsieh.25.PRL,Peterson.16.PRSA,Yan.18.PRL,Gaudenzi.18.NP,Aimet.25.NP,Chattopadhyay.25.RPP}. In macroscopic settings, a thermal bath assumption is often reasonable. However, when addressing thermodynamic cost of quantum processes at the nanoscale, this assumption frequently conflicts with experimental realities, including non-thermal environments~\cite{Langen.S.15,Alvarez.15.S,Degen.17.RMP,Yang.17.RPP,WangP.19.PRL,Abanin.19.RMP,Lupke.20.PRXQ,Jackson.21.NP,Kuffer.22.PRXQ,Kuffer.25.PRXQ} or complex and unknown environments~\cite{Wise.21.PRXQ,Martina.23.PS,Chen.19.NC,Spiecker.22.NP}. One proposed approach to circumvent this incompatibility is the introduction of environmental reference states~\cite{Vaccaro.11.PRSA,Barnett.13.E,Lostaglio.17.NJP,Timpanaro.20.PRL,Mondal.23.A}, which presupposes precise knowledge of the actual time-dependent environmental state. However, such knowledge is often unavailable in practice, especially for complex or engineered quantum systems where not all environmental and system degrees of freedom are experimentally accessible. Typically, only a subset of the system’s degrees of freedom are observable, and the prevalence of partially hidden dynamics is increasingly  recognized~\cite{Meer.22.PRX,Degunther.24.PRR,Blom.24.PNAS,Liu.23.PRAa,LiuJ.24.PRR}. This gap between theoretical assumptions and practical constraints calls for a more general thermodynamic framework capable of deriving Landauer-type information-cost trade-off relations~\footnote{We here explicitly refer to information-cost relations as those that establish fundamental bounds on thermodynamic cost in terms of the change in the system’s information content--quantified in the quantum regime by the von Neumann entropy~\cite{Lieb.02.NULL}.} when only partial information of the system is available.

%===============================================
\begin{figure*}[thb!]
 \centering
\includegraphics[width=1.5\columnwidth]{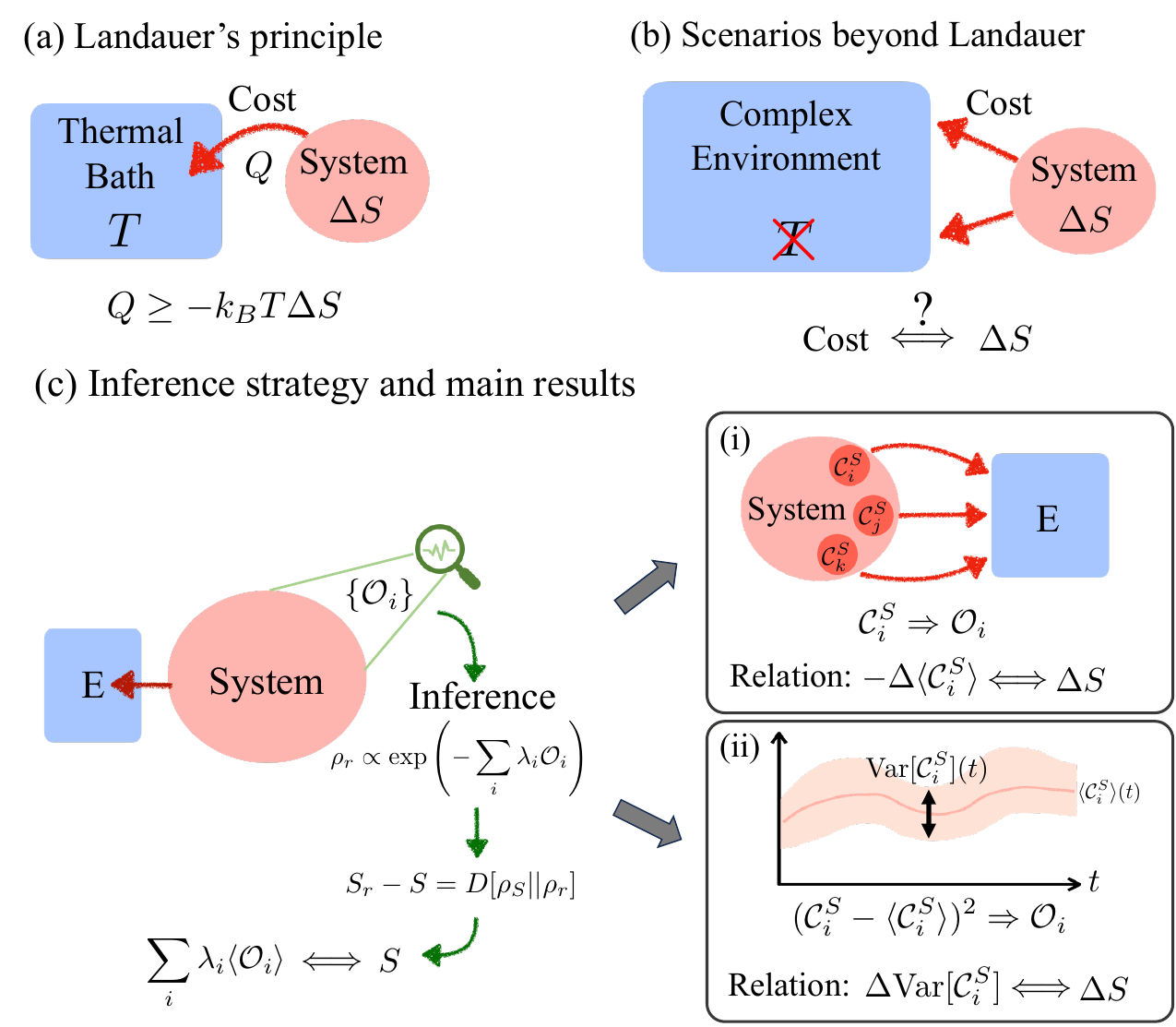} 
 \caption{Motivation and schematic of the study. (a) The celebrated Landauer's principle--a cornerstone information-cost relation--establishes a lower bound on heat cost $Q$ dissipated into a thermal bath at temperature $T$ in terms of the change in the system’s information content $\Delta S$ with $S=-\mathrm{Tr}[\rho_S\ln\rho_S]$. (b) However, at the nanoscale, quantum systems often couple to complex environments (E) such as non-thermal ones or those with limited experimental accessibility, where temperature becomes ill-defined. This raises a fundamental question: can an information-cost relation still be formulated under such conditions? (c) Here we address the challenge of limited experimental access to microscopic details by developing a thermodynamic inference framework that leverages available information about a set of system observables $\{\mathcal{O}_i\}$. We introduce a reference state $\rho_r$ [Eq. (\ref{eq:reference_state})] which is constructed via the maximum entropy principle. The difference between reference information content $S_r=-\mathrm{Tr}[\rho_r\ln\rho_r]$ and the system's actual information content $S$ is just the quantum relative entropy $D[\rho_S||\rho_r]=\mathrm{Tr}[\rho_S(\ln\rho_S-\ln\rho_r)]$ between the reference state $\rho_r$ and the actual state $\rho_S$. With this relation, we can connect expectation values $\langle \mathcal{O}_i\rangle$ (observations) with the system's actual information content $S$ [Eq. (\ref{eq:equality})]. We apply the approach to two scenarios that differ in the available system observations: (i) Quantum systems where only mean values $\langle\mathcal{C}_i^S\rangle$ of multiple conserved charges $\{\mathcal{C}_i^S\}$ (energy, particle number, etc) are accessible, for which we establish an information-cost relation between cost $-\Delta\langle\mathcal{C}_i^S\rangle(t)=\langle\mathcal{C}_i^S\rangle(0)-\langle\mathcal{C}_i^S\rangle(t)$ and system actual information content change $\Delta S(t)$ [Eq. (\ref{eq:Our_ex})], (ii) Quantum systems where both the mean values $\langle \mathcal{C}_i^S\rangle$ and the variances $\mathrm{Var}[\mathcal{C}_i^S]\equiv\langle (\mathcal{C}_i^S-\langle \mathcal{C}_i^S\rangle)^2\rangle$ are accessible, for which we establish an information-cost relation between fluctuation cost $\Delta\mathrm{Var}[\mathcal{C}_i^S](t)=\mathrm{Var}[\mathcal{C}_i^S](t)-\mathrm{Var}[\mathcal{C}_i^S](0)$ and system actual information content change $\Delta S(t)$ [Eq. (\ref{eq:var_eq})].}
\protect\label{fig:sketch}
\end{figure*}
%===============================================

Another key feature that distinguishes quantum thermodynamic processes from classical ones is their significant thermodynamic fluctuations. These fluctuations adhere to well-established fluctuation theorems~\cite{Esposito.09.RMP,Seifert.12.RPP} and are increasingly accessible to experimental measurement~\cite{Schweigler.17.N,Ciliberto.17.PRX,Impertro.24.PRL,Joshi.25.PRL}.
It is therefore natural to expect that controlled quantum processes should involve changes not only in mean values of energetic quantities such as heat and work but also in their higher-order fluctuations. This becomes apparent by considering an ideal qubit reset process which starts from a mixed state and ends in the ground state: here, both the averaged energy and the energy variance change substantially. Given the ubiquity and importance of thermodynamic fluctuations at the nanoscale, it is striking that existing formulations of quantum thermodynamic cost relations remain largely confined to changes in mean values of energetic quantities--mirroring their classical counterparts. Treating changes in observable fluctuations as higher-order thermodynamic costs--dubbed fluctuation costs hereafter--and incorporating them into a generalized framework of information-cost relations is also essential for a complete description of thermodynamic costs in the deep quantum regime.

%--------------------------------------------------------
\begin{table*}[thb!]
\centering
\caption{Summary of our main results. We consider system charges $\{\mathcal{C}_i^S\}$ (energy, particle number, etc) as the primary observables. Our first main result applies to arbitrary systems and requires only the mean values of these charges during the evolution. It establishes an upper bound on the system's charge loss $-\Delta \langle \mathcal{C}_i^S\rangle (t)\equiv\langle \mathcal{C}_i^S\rangle (0)-\langle \mathcal{C}_i^S\rangle (t)$ (thermodynamic mean cost), thereby complementing a generalized Landauer's principle~\cite{Lostaglio.17.NJP} that provides a lower bound on the corresponding charge gain $\Delta \langle \mathcal{C}_i^B\rangle (t)$ in the bath. The second main result applies to arbitrary systems where both the mean values and variances of the charges are accessible. It yields a lower bound on the change in variance $\Delta \mathrm{Var}[\mathcal{C}_i^S](t)\equiv\mathrm{Var}[\mathcal{C}_i^S](t)-\mathrm{Var}[\mathcal{C}_i^S](0)$ (thermodynamic fluctuation cost) for which no prior bounds exist.}
\setlength{\extrarowheight}{5pt}
\begin{tabularx}{\textwidth}{lcc}
\hline
\hline
Available system information  &  \hspace{1cm}Information-cost relation & \hspace{1cm}Existing result \\ 
\hline
Mean value $\langle \mathcal{C}_i^S\rangle(t)$ (Sec. \ref{sec:2}) & \hspace{0.5cm}-$\sum_i\lambda_i(0)\Delta \langle \mathcal{C}_i^S\rangle(t)\le\mathcal{U}_M(t)$ [Eq. (\ref{eq:Our_ex})] & \hspace{0.8cm}$\sum_i\mu_i\Delta\langle\mathcal{C}_i^B\rangle(t)\ge -\Delta S(t)$ \cite{Lostaglio.17.NJP} \\
\hline
Mean value $\langle \mathcal{C}_i^S\rangle(t)$ and variance $\mathrm{Var}[\mathcal{C}_i^S](t)$ (Sec. \ref{sec:3}) & \hspace{0.5cm} $\Delta\mathrm{Var}[\mathcal{C}_i^S](t)\ge \mathcal{L}_v(t)$ [Eq. (\ref{eq:var_eq})] & \hspace{0.8cm}NA \\
\hline
\hline
\end{tabularx}
\label{table1}
\end{table*}
%--------------------------------------------------------
Motivated by these practical needs to extend current frameworks for deriving information-cost relations, in this study, we develop a thermodynamic inference approach based on the maximum entropy principle~\cite{Jaynes.57.PR} (see Fig. \ref{fig:sketch} for a sketch). This approach enables us to address aforementioned needs within a single framework. Unlike existing efforts that apply inference strategies to infer environmental reference states~\cite{Vaccaro.11.PRSA,Barnett.13.E,Lostaglio.17.NJP,Timpanaro.20.PRL,Mondal.23.A}, we use the maximum entropy principle to construct a reference system state that is consistent with a limited set of system observations--be it the mean values of system observables or their higher-order fluctuations--while remaining maximally unbiased with respect to unknown information. We note that the difference in information content as quantified by the von Neumann entropy~\cite{Lieb.02.NULL} between the actual system state and this reference state is precisely given by their quantum relative entropy, which enables us to connect observations with system's actual information content. This connection provides a foundation for establishing information-cost trade-off relations that bear no existing counterparts. By construction, our approach is inherently agnostic to the environment, relying solely on incomplete system-level information and naturally accommodates higher-order fluctuations.

To demonstrate the utility of our approach, we apply it to two  scenarios of broad interest (see Fig. \ref{fig:sketch}) to derive information-cost relations that complement and extend the scope of the LP. Both scenarios concern quantum systems with charges--such as energy and particle number--as observables, but differ in the available system observations. In the first scenario, we consider systems with multiple charges that have received
increasing attention in quantum thermodynamics~(see a recent comprehensive review~\cite{Majidy.23.NRP} and references therein) and assume that only the mean values of these charges are accessible as experimental observations. In a manner analogous to the treatment of heat dissipation in the LP, we identify system charge loss during arbitrary evolution as the corresponding thermodynamic cost. We derive a general
information-cost relation that establishes a fundamental upper bound on the thermodynamic cost in terms of the change in the system’s actual information content. If more information about the state of the coupled environment is accessible, a generalized Landauer lower bound~\cite{Lostaglio.17.NJP} which assumes a generalized Gibbsian state~\cite{Jaynes.57.PR} for the environment becomes applicable. In this special case, our upper bound can be combined with it to form a two-sided bound on the thermodynamic cost at finite times; In the quasi-static limit, we demonstrate that the lower and upper bounds converge to the same value.

In the second scenario, we consider systems where the variances of system charges are further accessible as experimental observations. We introduce the concept of thermodynamic fluctuation cost which is defined as the change in the variance during arbitrary evolution. We derive a general information-cost relation that provides a lower bound on this fluctuation cost in terms of the change in the system’s actual information content. This result generalizes the rationale behind the LP by incorporating changes in higher-order fluctuations and treating them as a form of higher-order thermodynamic cost. We note that the celebrated thermodynamic uncertainty relation (TUR)~\cite{Barato.15.PRL,Gingrich.16.PRL,Horowitz.19.NP,Liu.19.PRE} establishes a link between instantaneous variance and total entropy production and is often viewed as a refinement of the LP to variance. However, the LP fundamentally relates system energy change (as heat) to system entropy change. In this sense, our information–cost relation in the second scenario aligns more closely with the spirit of the LP than with the TUR, as it connects the change in variance to the change in system entropy. We demonstrate the validity and practical relevance of these two information-cost relations through detailed numerical simulations of three systems: a coupled-qubit system exchanging energy and excitations, a driven qubit implementing information erasure, and a driven double quantum dot that can operate as an inelastic heat engine. 

The structure of this paper is as follows. In Sec. \ref{sec:1}, we present the core of our thermodynamic inference framework. This includes a general form of the system reference state, constructed via the maximum entropy principle under constraints from available system information, and an equality that links known system observables to the actual information content of the system. In Sec. \ref{sec:2}, we consider systems with multiple conserved charges where only mean charge values are accessible. We derive an information-cost relation that provides a general upper bound on the thermodynamic cost and compare it with an existing lower bound--within its regime of validity--both at finite times and in the quasi-static limit. The upper bound is further validated using a minimal model with two conserved charges. In Sec. \ref{sec:3}, we extend the analysis to scenarios where the knowledge of fluctuations becomes further available. We derive an information-cost relation yielding a general lower bound on the change in observable fluctuations during a process, and verify it numerically using a driven qubit model and a driven double quantum dot system. We conclude in Sec. \ref{sec:4} with final remarks. For clarity, we summarize our main results in Table~\ref{table1}.

\section{Thermodynamic inference from the maximum entropy principle}\label{sec:1}
In the most general case, only partial system's degrees of freedom are observables. Let $\{\mathcal{O}_i\}$ with $i=0,\cdots,K-1$ be such a set of system observables whose expectation values $\{\langle \mathcal{O}_i\rangle(t)\}$ can be experimentally determined during a finite-time process. This set of expectation values defines the limited knowledge available about the system at time $t$. Based on this available information, the most unbiased inference for the system state is given by the maximum entropy state \cite{Jaynes.57.PR} that maximizes the system's von Neumann entropy (i.e., the system's information content~\cite{Lieb.02.NULL}), while ensuring that it satisfies given dynamical constraints set by measured expectation values $\{\langle \mathcal{O}_i\rangle(t)\}$. We set $k_B=1$ and $\hbar=1$ hereafter.

Using the method of Lagrange multipliers where the available information about observables is used as dynamical constraints, we can construct a Lagrangian functional $\mathcal{L}[\rho]$ about a system state $\rho$ (time dependence is suppressed for simplicity),
\begin{equation}
    \mathcal{L}[\rho] ~=~ -\mathrm{Tr}[\rho\ln\rho]+\eta (\mathrm{Tr}[\rho]-1)+\sum_{i=0}\lambda_i\left[\mathrm{Tr}(\rho \mathcal{O}_i)-\langle \mathcal{O}_i\rangle \right].
\end{equation}
Here, $\eta$ is a Lagrange multiplier that determines the normalization factor. $\{\lambda_i\}$ are Lagrange multipliers associated with the dynamical constraints $\mathrm{Tr}(\rho \mathcal{O}_i)=\langle \mathcal{O}_i\rangle$. Setting the functional derivative with respect to $\rho$ to zero, $\partial \mathcal{L}/\partial \rho=0$, we receive the maximum entropy state 
\begin{equation}\label{eq:reference_state}
    \rho_r(t)~=~\frac{1}{Z_r(t)}\exp\left(-\sum_{i=0}\lambda_i(t)\mathcal{O}_i\right).
\end{equation}
Here, $Z_r(t)=\mathrm{Tr}\left[\exp\left(-\sum_{i=0}^{K-1}\lambda_i(t)\mathcal{O}_i\right)\right]$ is the normalization factor, and the Lagrange multipliers $\{\lambda_i\}$ ensure that the maximum entropy state matches the observations, 
\begin{equation}\label{eq:constraint}
-\frac{\partial \ln Z_r(t)}{\partial \lambda_i(t)}~=~\mathrm{Tr}[\rho_r(t)\mathcal{O}_i]~=~\langle \mathcal{O}_i\rangle(t).
\end{equation}
Moreover, the Lagrange multipliers $\{\lambda_i(t)\}$ represent effective potentials associated with the observables $\{\mathcal{O}_i\}$ from a thermodynamic perspective, providing the necessary dimensional parameters to construct information–cost relations. Theoretically, the observation $\langle \mathcal{O}_i\rangle(t)=\mathrm{Tr}[\rho_S(t)\mathcal{O}_i]$ is defined as the expectation value of the associated observable $\mathcal{O}_i$ with respect to the actual system state $\rho_S(t)$. However, we emphasize that this value can be obtained experimentally from an ensemble of measurements on the observable $\mathcal{O}_i$, without actually requiring the knowledge of the actual system state $\rho_S$. That is, determining the explicit form of the system reference state in Eq. (\ref{eq:reference_state}) requires only knowledge of the set of experimentally accessible system operators $\{\mathcal{O}_i\}$, whose ensemble averages $\{\langle \mathcal{O}_i\rangle(t)\}$ represent the experimental observations and are used to determine the associated Lagrange multipliers according to Eq. (\ref{eq:constraint}).

Importantly, the maximum entropy state in Eq. (\ref{eq:reference_state}) represents the least biased or most uncertain state consistent with the available information--namely, the given observations of the specified observables. We emphasize that this maximum entropy state serves only as an inferred reference for nonequilibrium quantum systems and generally differs from the actual system state $\rho_S$. In what follows, we interpret this maximum entropy state as a reference state for nonequilibrium quantum systems. 

Considering the information content of the reference state $\rho_r$~\cite{Lieb.02.NULL}, we find that its deviation from the actual information content of the system is given by
\begin{equation}\label{eq:s_devi}
   S_r(t)-S(t)~=~D[\rho_S(t)||\rho_r(t)].
\end{equation}
Here, $S(t)=-\mathrm{Tr}[\rho_S(t)\ln \rho_S(t)]$ and $S_r(t)=-\mathrm{Tr}[\rho_r(t)\ln \rho_r(t)]$ denote the information content of the actual state and the reference state, respectively, and $D[\rho_S(t)||\rho_r(t)]=\mathrm{Tr}[\rho_S(t)(\ln\rho_S(t)-\ln\rho_r(t))]$ is the quantum relative entropy between $\rho_S(t)$ and $\rho_r(t)$. To get the above relation, we have utilized the identity $\mathrm{Tr}[\rho_{r}(t)\ln\rho_r(t)]=\mathrm{Tr}[\rho_{S}(t)\ln \rho_r(t)]$, which follows from the constraint in Eq. (\ref{eq:constraint}). By definition, the quantum relative entropy is non-negative. Therefore, Eq. (\ref{eq:s_devi}) naturally implies that the uncertainty associated with the reference state is generally greater than that of the actual system state, reflecting the increased uncertainty inherent in the maximum entropy inference based on limited knowledge.

Using Eq. (\ref{eq:reference_state}) to rewrite $S_r(t)$ in Eq. (\ref{eq:s_devi}), we immediately get an equality that connects observations $\{\langle \mathcal{O}_i\rangle(t)\}$ with the actual system information content $S(t)$~\cite{Balian.07.NULL},
\begin{equation}\label{eq:equality}
    \sum_{i=0}\lambda_i(t)\langle\mathcal{O}_i\rangle(t)-S(t)~=~-\ln Z_r(t)+D[\rho_S(t)||\rho_r(t)].
\end{equation}
We note that the left-hand-side of Eq. (\ref{eq:equality}) is just the free entropy $\widetilde{F}_S(\rho_S)$ introduced in Ref. \cite{Guryanova.16.NC}. Hence, the above equation can also be rewritten as $\widetilde{F}_S(\rho_S)-\widetilde{F}_S(\rho_r)=D[\rho_S||\rho_r]$. In the special case where the system internal energy $E=\langle H\rangle$ represents the only observation, the reference state $\rho_r$ takes the form of a Gibbsian state~\cite{Jaynes.57.PR}, with the corresponding Lagrange multiplier $\lambda$ have the dimension of inverse temperature. In this limit, Eq. (\ref{eq:equality}) reduces to a generalized notion of nonequilibrium free energy, $\mathcal{F}=\lambda^{-1}[-\ln Z_r+D(\rho_S||\rho_r)]=E-\lambda^{-1}S$ which applies to both thermal and non-thermal processes \cite{Liu.23.PRAa,LiuJ.24.PRR}. 

We emphasize that the equality in Eq. (\ref{eq:equality}) holds for arbitrary nonequilibrium thermodynamic and information-processing processes, establishing an exact relationship between the system's actual information content and experimentally accessible observables through the reference state $\rho_r$. In what follows, we demonstrate how this equality enables the derivation of nontrivial information-cost relations that yield general bounds on thermodynamic cost. These relations reveal fundamental trade-offs between the consumption of system's observables as resources and changes in system's actual information content during nonequilibrium processes. Such results are unattainable with existing methodologies based on the second law of thermodynamics, thereby complementing and extending the scope of the LP.

In view of typical experimental capacities, we explicitly focus on system observables such as energy and particle number as the set of accessible observables in the following. For later convenience, we refer to them as charges $\{\mathcal{C}_i^S\}$ as they represent conserved quantities in closed setups. We emphasize that quantum systems may possess multiple such charges. Below, we distinguish and analyze two scenarios separately [See Fig. \ref{fig:sketch} (c)]: (i) In the first scenario, we assume that only first-order observations, namely, mean values of charges, are available. Here, the operators $\{\mathcal{O}_i\}$ in Eq. (\ref{eq:reference_state}) correspond directly to the charge set $\{\mathcal{C}_i^S\}$. (ii) In the second scenario, we further assume that second-order observations (variances) of the charges are accessible. In this case, the operator $\mathcal{O}_i$ in Eq. (\ref{eq:reference_state}) becomes $(\mathcal{C}_i^S-\langle\mathcal{C}_i^S\rangle)^2$ whose ensemble average just gives the variance.

%================================================================
\section{Quantum systems knowing mean values}\label{sec:2}
In this section, we focus on open quantum systems with multiple charges $\{\mathcal{C}_i^S\}$~\footnote{We emphasize that the conservation of charges in open systems is defined in the global sense, that is, they are preserved in the composite system consisting of the system of interest and its environment, while allowing exchange between them.}. These charges may include energy, angular momentum, and other relevant observables. During nonequilibrium processes, such systems can exchange charges with their environments. Noting that quantum thermodynamic systems with multiple charges have attracted growing theoretical and experimental interest~\cite{Majidy.23.NRP}, making it essential to understand thermodynamic costs associated with driving them~\cite{Vaccaro.11.PRSA,Barnett.13.E,Lostaglio.17.NJP,Halpern.18.JPA,Mondal.23.A}. Here, given the mean values of charges as the available knowledge about the system, we establish a general information-cost relation from the inference perspective introduced above. This relation imposes new constraints on the thermodynamic cost in nonequilibrium processes involving multiple charges.

\subsection{Information-cost relation}
When the dynamical constraints are given by the mean values of the charges $C_i^S\equiv\langle\mathcal{C}_i^S\rangle$, the corresponding reference state in Eq. (\ref{eq:reference_state}) takes the specific form of a generalized Gibbsian state~\cite{Jaynes.57.PR,Lostaglio.17.NJP,Halpern.16.NC,Guryanova.16.NC},
\begin{equation}
    \rho_r(t)~=~\frac{1}{Z_r(t)}\exp\left(-\sum_{i}\lambda_i(t)\mathcal{C}_i^S\right).
\end{equation}
That is, we replace a generic operator $\mathcal{O}_i$ in the general form of Eq. (\ref{eq:reference_state}) with the system charge $\mathcal{C}_i^S$ in this scenario. A similar substitution applies to Eq. (\ref{eq:equality}). After this replacement, we use Eq. (\ref{eq:equality}) to express the change in system's actual information content over the time interval $[0,t]$ as
\begin{equation}\label{eq:multi_equality}
    \Delta S(t)~=~\sum_{i=0} \lambda_i(0) \Delta C_i^S(t)+\mathcal{R}(t)-D[\rho_{S}(t)||\rho_r(t)].
\end{equation}
Here, we have denoted the change $\Delta A(t)=A(t)-A(0)$ for an arbitrary quantity $A$. In analogy with heat dissipation which represents system energy loss in forms of heat considered in the LP, we interpret the loss of averaged system charge $-\Delta C_i^S(t)=C_i^S(0)-C_i^S(t)$ as the thermodynamic cost associated with the charge $\mathcal{C}_i^S$ during the time interval $[0,t]$. We have also defined $\mathcal{R}(t)=\sum_{i=0} C_i^S(t) \Delta \lambda_i(t)+\ln\left( {Z_r(t)}/{Z_r(0)}\right)+ \Delta S^{in}$; where $\Delta S^{in}=D[\rho_S(0)||\rho_r(0)]=\mathrm{Tr}[\rho_{S}(0) \ln\rho_S(0)]-\mathrm{Tr}[\rho_r(0) \ln\rho_r(0)]$ is the initial entropy difference derived using Eq. (\ref{eq:constraint}), $\Delta \lambda_i(t)=\lambda_i(t)-\lambda_i(0)$. We note that $\mathcal{R}(t)$ vanishes only when the system initially resides in a generalized Gibbsian ensemble and remains in it throughout the process, which implies $\Delta \lambda_i(t)=0$, $Z_r(t)=Z_r(0)$ and $\Delta S^{in}=0$. Hence, $\mathcal{R}(t)$ quantifies the degree of nonequilibrium in quantum processes relative to the reference generalized Gibbsian state. For our convention, we denote the component $C_0^S = E_S$ as the system internal energy with the corresponding charge $\mathcal{C}_0^S=H_S$.

Due to the non-negativity of quantum relative entropy, we can turn the above equality into an inequality  
\begin{equation}\label{eq:Our_ex}
-\sum_{i=0}\lambda_i(0)\Delta C_i^S(t)~\le~\mathcal{U}_M(t).
\end{equation}
Here, we have defined $\mathcal{U}_M(t)=\mathcal{R}(t)-\Delta S(t)$ which incorporates the change in the system's actual information content $\Delta S(t)$, a desire feature we seek for extending the LP. Remarkably, the above inequality establishes a thermodynamic upper bound on the dimensionless weighted thermodynamic cost $-\sum_i\lambda_i(0)\Delta C_i^S(t)$ dissipated from the system. Eq. (\ref{eq:Our_ex}) represents our first main result on information-cost relations. We emphasize that Eq. (\ref{eq:Our_ex}) can also be reformulated as a thermodynamic upper bound on an individual component of thermodynamic cost $-\lambda_i(0)\Delta C_i^S(t)$. To evaluate $\mathcal{U}_M(t)$ and thereby verify the whole information-cost relation Eq. (\ref{eq:Our_ex}) for a given dynamical process, one requires the mean values of charges at the initial and the final times to determine the corresponding Lagrange multipliers based on Eq. (\ref{eq:constraint}) and, consequently, the value of $\mathcal{R}(t)$.

We remark that this information-cost relation Eq. (\ref{eq:Our_ex}) remains agnostic with respect to environmental details and initial system-environment states, owing to the solely system-dependent nature of the reference state within our inference framework. Hence, Eq. (\ref{eq:Our_ex}) is broadly applicable to general nonequilibrium processes in quantum systems coupled to diverse environments. When the system internal energy is the only accessible observable, Eq. (\ref{eq:Our_ex}) reduces to a recently reported result~\cite{Liu.23.PRAa,LiuJ.24.PRR}.

\subsection{Contrasting a generalized Landauer lower bound}
\subsubsection{Reformulation and comparison}
In a special scenario where the bath is known {\it a priori} to be initialized in a generalized Gibbsian state $\gamma_B=\exp[-\sum_{i=0}\mu_i\mathcal{C}_i^B]/Z_B$ with the partition function $Z_B=\mathrm{Tr}\left(\exp[-\sum_{i=0}\mu_i\mathcal{C}_i^B]\right)$ and bath thermodynamic potentials $\{\mu_i\}$ conjugate to bath charges $\{\mathcal{C}_i^B\}$, Ref. \cite{Lostaglio.17.NJP} generalized the LP to obtain a thermodynamic lower bound on the cost for quantum systems with multiple charges
\begin{equation}\label{eq:LB_NJP}
\sum_{i=0} \mu_{i}\Delta C_{i}^{B}(t)~\ge~-\Delta S(t). 
\end{equation}
Here, $C_i^B(t)=\mathrm{Tr}[\rho_B(t)\mathcal{C}_i^B]$ denotes the expectation value of the bath charge $\mathcal{C}_i^B$ with respect to bath state at time $t$. This implies that verifying Eq. (\ref{eq:LB_NJP}) requires tracking the evolution of the environment--a task that is challenging from both theoretical and experimental perspectives. We note that the results in Refs. \cite{Vaccaro.11.PRSA,Barnett.13.E} represent special cases of Eq. (\ref{eq:LB_NJP}) for systems with just two conserved charges (specifically, energy and spin angular momentum). When only energy is exchanged, the above inequality reduces to the standard LP by identifying the bath energy gain as heat dissipation from the system.

To enable a direct comparison between Eqs. (\ref{eq:Our_ex}) and (\ref{eq:LB_NJP}) within the validity regime of the latter, we need to establish the relation between $\Delta C_{i}^{S}(t)$ as used in Eq. (\ref{eq:Our_ex}) and $\Delta C_{i}^{B}(t)$ from Eq. (\ref{eq:LB_NJP}). We limit our analysis to the regime of weak system–environment coupling where such a relation is well-defined. We first note that for autonomous systems devoid of external controls or agents that can exchange charges with the system, charge conservation implies $\Delta C_{i}^{S}(t)+\Delta C_{i}^{B}(t)=0$ and hence $\Delta C_{i}^{S}(t)=-\Delta C_{i}^{B}(t)$. In the presence of external agents, we generally expect $\Delta C_{i}^{S}(t)+\Delta C_{i}^{B}(t)=W_i^S(t)$ with $W_i^S(t)$ being a work-like contribution performed on the system~\cite{Guryanova.16.NC}. By analogy with the conventional first law of thermodynamics for energy $\Delta E_S=-Q_0^S+W_0^S$, we can re-express the charge conservation as $\Delta C_i^S(t)=-Q_i^S(t)+W_i^S(t)$ where we identify $Q_i^S(t)=\Delta C_{i}^{B}(t)$ as the heat-like contribution flowing into the bath and leading to a gain in bath charge. This decomposition is physically motivated by the fact that the bath exchanges only heat-like quantities with the system. If the system is initially thermalized with the bath in a generalized Gibbsian state, we may set $\lambda_i(0)=\mu_i$ in Eqs. (\ref{eq:Our_ex}) and (\ref{eq:LB_NJP}). Then the above decomposition allows us to rewrite the generalized LP [cf. Eq. (\ref{eq:LB_NJP})] as
\begin{equation}\label{eq:LB_NJP_2}
-\sum_{i=0} \lambda_i(0)\Delta C_{i} ^{S}(t)~\ge~-\Delta S(t)-\sum_{i=0}\lambda_i(0)W_i^S(t).
\end{equation}
This form makes it clear that the generalized LP provides a thermodynamic lower bound on the dimensionless thermodynamic cost $-\sum_i\lambda_i(0)\Delta C_i^S(t)$ considered in Eq. (\ref{eq:Our_ex}). Therefore, our result Eq. (\ref{eq:Our_ex}) complements the generalized LP Eq. (\ref{eq:LB_NJP}) by supplying a corresponding upper bound valid at finite times. Importantly, beyond this special scenario of a bath in a generalized Gibbsian state, Eq. (\ref{eq:LB_NJP}) becomes inapplicable as $\{\mu_i\}$ are ill-defined, whereas our result Eq. (\ref{eq:Our_ex}) remains valid due to its independence from environmental details.

\subsubsection{Quasi-static limit}
For finite-time quantum processes, neither Eq. (\ref{eq:Our_ex}) nor Eq. (\ref{eq:LB_NJP_2}) attains equality; instead, they just complement each other. However, we will show in the below that both inequalities saturate to the same bound in the quasi-static limit.

Under quasi-static conditions, the generalized LP in Eq. (\ref{eq:LB_NJP_2}) saturates to an equality~\cite{Lostaglio.17.NJP}. Regarding the upper bound in Eq. (\ref{eq:Our_ex}), we first notice that if the system is initially thermalized to {a generalized Gibbsian state with affinities fixed by the bath ones, it remains in that state in the course of a quasi-static process, rendering $D[\rho_S(t)||\rho_r(t)]$ vanishing. Consequently, Eq. (\ref{eq:Our_ex}) should also becomes an equality, as followed from Eq. (\ref{eq:multi_equality}). This implies that the upper bound [cf. Eq. (\ref{eq:Our_ex})] and the lower bound [cf. Eq. (\ref{eq:LB_NJP_2})] on the same quantity $-\sum_{i=0}\lambda_i(0)\Delta C_i^S(t)$ converge to the same value in the quasi-static limit, providing a consistent characterization of thermodynamic cost in reversible processes.

To provide a deeper analysis of this convergence in the quasi-static limit, we adopt a detailed yet idealized quasi-static protocol~\cite{Lostaglio.17.NJP} that saturates Eq. (\ref{eq:LB_NJP}) and tailor it to examine Eq. (\ref{eq:Our_ex}). In this specially designed protocol, the system has just two charges with $\mathcal{C}_0^S=H_S(t)$ and $\mathcal{C}_1^S(t)$, both of which may be time-dependent due to the presence of external agents. Likewise, the bath has two corresponding charges $\mathcal{C}_0^B=H_B$ and $\mathcal{C}_1^B$ that can be exchanged with the system. We assume the bath stays in a generalized Gibbsian state $\gamma_B=\exp[-\beta_BH_B-\mu_1\mathcal{C}_1^B]/\mathrm{Tr}[e^{-\beta_BH_B-\mu_1\mathcal{C}_1^B}]$ with $\mu_0=\beta_B$ denoting the usual inverse temperature of the bath. During the quasi-static process, the system remains in a generalized Gibbsian state determined by the instantaneous observables $H_S(t)$ and $\mathcal{C}_1^S(t)$ such that the inferred reference state simply reads $\rho_r(t)=\exp[-\beta_BH_S(t)-\mu_1\mathcal{C}_1^S(t)]/Z_r(t)$; Here, we have used the relations $\lambda_0(0)=\lambda_0(t)=\beta_B$ and $\lambda_1(0)=\lambda_1(t)=\mu_1$ as the system's generalized Gibbsian state is fixed by the bath under quasi-static conditions. In this special case, $\mathcal{R}(t)$ in Eq. (\ref{eq:Our_ex}) simplifies to $\ln\left(Z_r(t)/Z_r(0)\right)$. It remains to verify whether $\sum_{i=0}\lambda_{i}(0)\Delta C_i^S(t)+ \ln\left(Z_r(t)/Z_r(0)\right)$ equals $\Delta S(t)$ during this quasi-static protocol.

To specify the quasi-static protocol in detail, we first outline some necessary assumptions. For systems possessing only two charges, the commutation relations $[H_B,\mathcal{C}_1^B]=[H_S,\mathcal{C}_1^S]=0$ hold. 
%This allows the bath state $\gamma_B$ to be formally factorized, effectively representing the bath as two independent sub-baths: an energy bath and a `$\mathcal{C}_1$-bath'. We assume that the system can be coupled to each of these two sub-baths separately. 
For simplicity, we assume that both $H_S$ and $\mathcal{C}_1^S$ share a common two-level eigenbasis $\{|0\rangle,|1\rangle\}$ as they commute. It is convenient for us to work in a joint state space that is spanned by four basis states with an order $\{|00\rangle,|01\rangle,|10\rangle,|11\rangle\}$, where $|nm\rangle\equiv |n\rangle\otimes |m\rangle$ ($n,m=0,1$) with $|n\rangle$ and $|m\rangle$ denoting eigenstates of $H_S$ and $\mathcal{C}_1^S$, respectively. In this joint space, all operators are represented in tensor product form: for instance, the charges take the form $H_S\otimes \mathrm{I}$ and $\mathrm{I}\otimes \mathcal{C}_1^S$, where $\mathrm{I}$ is the $2\times 2$ identity matrix. With this in mind, we will simplify the notations by suppressing explicit tensor product symbols in what follows. These states are initially degenerate in both $H_S(0)$ and $\mathcal{C}_1^S(0)$, which can be achieved by setting the initial eigenvalues of $H_S(0)$ and $\mathcal{C}_1^S(0)$ associated with eigenstates $\{|0\rangle, |1\rangle\}$ to zero. With these preliminaries, the detailed quasi-static protocol tailored to our result in Eq. (\ref{eq:Our_ex}) is implemented as follows (Our protocol greatly refines that of Ref. \cite{Lostaglio.17.NJP}).

(i) We initially prepare the system in the state
\begin{equation}
    \rho_S(0)~=~\frac{1}{4}\left(|0\rangle\langle 0|+|1\rangle\langle 1|\right)\otimes\left(|0\rangle\langle 0|+|1\rangle\langle 1|\right).
\end{equation}
$\rho_S(0)$ is the maximally mixed state in the joint space. Noting that the eigenvalues of $H_S(0)$ and $\mathcal{C}_1^S(0)$ are zero, this state can be mapped onto a thermal state with respect to trivial charges $H_S(0)=0$ and $\mathcal{C}_1^S(0)=0$. We accordingly infer that the Gibbsian reference state satisfies $\rho_r(0)=\rho_S(0)$, with $Z_r(0)=4$ and $\Delta S^{in}=0$ in Eq. (\ref{eq:Our_ex}).

(ii) We then adjust the Hamiltonian to gradually shift the energy eigenvalue of the energy eigenstate $|1\rangle$ from $0$ to $\epsilon$ through a sequence of $n\to\infty$ steps over a long time interval $[0,t_1]$. Under quasi-static controls, the system remains in a thermal state with respect to the instantaneous Hamiltonian $H_S(t)=\epsilon'|1\rangle\langle 1|$ and $\mathcal{C}_1^S(t)=0$ where $t\in[0,t_1]$,$\epsilon'\in[0,\epsilon]$; We remind that $H_S(t)$ is represented by a four-dimensional diagonal matrix, $H_S(t)=\mathrm{diag}\{0,0,\epsilon',\epsilon'\}$, in the joint space. During this step, the system state is given by
\bea\label{eq:srho_1}
   \rho_S(t)~=~\rho_r(t) &=& \frac{1}{2(1+e^{-\beta_B\epsilon'})}\left(|0\rangle\langle 0|+e^{-\beta_B\epsilon'}|1\rangle\langle1|\right)\nonumber\\
   &&\otimes \left(|0\rangle\langle 0|+|1\rangle\langle 1|\right).
\eea
Here, $\beta_B$ is the inverse temperature of the energy bath. In the limit of $\epsilon=0$, $\rho_S(t)$ in Eq. (\ref{eq:srho_1}) reduces to $\rho_S(0)$.

At the end of this step, we have $\rho_S(t_1)=\rho_r(t_1)=\frac{1}{2(1+e^{-\beta_B\epsilon})}\left(|0\rangle\langle 0|+e^{-\beta_B\epsilon}|1\rangle\langle1|\right)\otimes \left(|0\rangle\langle 0|+|1\rangle\langle 1|\right)$ such that $Z_r(t_1)=2(1+e^{-\beta_B\epsilon})$. The work performed on the system to tune the Hamiltonian during this step is~\cite{Lostaglio.17.NJP}
\begin{equation}
    W_0^S[0,t_1]~=~\int_0^\epsilon \frac{e^{-\beta_B\epsilon'}}{1 + e^{-\beta_B\epsilon'}} d\epsilon'~=~\frac{1}{\beta_B}\ln\left(\frac{2}{1+e^{-\beta_B \epsilon}}\right).
\end{equation}
Hereafter, we use notation $A[t',t'']$ to denote the change of $A$ over the time interval $[t',t'']$. It can be readily verified that the work-like contribution $W_0^S[0,t_1]$ during the time interval $[0,t_1]$ exactly cancels the term $\mathcal{R}(t)$ in Eq. (\ref{eq:Our_ex}) since $\beta_BW_0^S[0,t_1]+\ln(Z_r(t_1)/Z_r(0))=0$ during this step which just reflects the saturation of the second law of thermodynamics in the quasi-static limit. During this step, the energy change reads $\Delta E_S[0,t_1]=\mathrm{Tr}[\rho_S(t_1)H_S(t_1)]-0=\epsilon e^{-\beta_B\epsilon}/(1+e^{-\beta_B \epsilon})$, while the expectation value of the other charge remains unchanged with $\Delta C_1^S[0,t_1]=0$ by noting that eigenvalues of $\mathcal{C}_1^S$ remain zero. Hence
the heat-like contribution $ Q_0^S[0,t_1]\equiv W_0^S[0,t_1]-\Delta E_S[0,t_1]$ during the time interval $[0,t_1]$ reads
\begin{equation}
    Q_0^S[0,t_1]~=~\frac{1}{\beta_B}\ln\left(\frac{2}{1+e^{-\beta_B \epsilon}}\right)-\frac{\epsilon e^{-\beta_B\epsilon}}{1+e^{-\beta_B \epsilon}}.
\end{equation}

(iii) We implement a unitary SWAP operation that exchanges the states $|01\rangle$ and $|10\rangle$ during a time interval $[t_1, t_2]$. The system state $\rho_S(t_1)\to\rho_S(t_2)$, where
\bea\label{eq:srho_t2}
    \rho_S(t_2) &=& \frac{1}{Z_r(t_1)}\left(|0\rangle\langle 0|+|1\rangle\langle 1|\right)\otimes\left(|0\rangle\langle 0|+e^{-\beta_B\epsilon}|1\rangle\langle 1|\right)\nonumber\\
    &=& \frac{1}{Z_r(t_2)}\left(|0\rangle\langle 0|+|1\rangle\langle 1|\right)\otimes\left(|0\rangle\langle 0|+e^{-\mu_1\eta}|1\rangle\langle 1|\right).\nonumber\\
\eea
Here, we have set $\eta\mu_1=\beta_B\epsilon$ to obtain the second line, which yields $Z_r(t_2)=2(1+e^{-\mu_1\eta})=Z_r(t_1)$. Comparing states $\rho_S(t_1)$ and $\rho_S(t_2)$, we see that this unitary SWAP operation amounts to changing the energy eigenvalue of $|1\rangle$ from $\epsilon$ to 0, and simultaneously changing the charge eigenvalue of $|1\rangle$ from 0 to $\eta=\beta_B\epsilon/\mu_1$. By doing so, $\rho_S(t_2)$ remains a thermal state with respect to Hamiltonian $H_S(t_2)=0$ and $\mathcal{C}_1^S(t_2)=\eta|1\rangle\langle 1|$. Because the eigenvalues have been modified during the unitary step, we have quantities changes $\Delta E_S[t_1,t_2]=-\epsilon e^{-\beta_B\epsilon}/(1+e^{-\beta_B\epsilon})$ and $\Delta C_1^S[t_1,t_2]=\eta e^{-\mu_1\eta}/(1+e^{-\mu_1\eta})$ during the time interval $[t_1,t_2]$.

(iv) Lastly, we gradually increase the eigenvalue $\eta$ of $\mathcal{C}_1^S$ to infinity during a time interval $[t_2,t_f]$ (where $t_f\to\infty$ ideally, but we retain the notation for convenience). According to Eq. (\ref{eq:srho_t2}), the final state is
\begin{equation}
    \rho_S(t_f)=\rho_r(t_f)=\frac{1}{2}\left(|00\rangle\langle 00|+|10\rangle\langle 10|\right),
\end{equation}
from which we have $Z_r(t_f)=2$. To complete the step, an amount of work-like contribution is needed to exert on the system to raise the eigenvalue of the charge $\mathcal{C}_1^S$,
\begin{equation}
W_1^S[t_2,t_f]~=~\int_{\eta}^{\infty} \frac{e^{-\mu_1\eta'}}{1+e^{-\mu_1\eta'}}d\eta'~=~\frac{1}{\mu_1}\ln(1+e^{-\mu_1\eta}).
\end{equation}
During this step we also find that the work-like contribution $W_1^S[t_2,t_f]$ cancels the term $\mathcal{R}(t)$ in Eq. (\ref{eq:Our_ex}) as $\mu_1W_1^S[t_2,t_f]+\ln(Z_r(t_f)/Z_r(t_2))=0$. The remaining heat-like contribution $Q_1^S[t_2,t_f]=W_1^S[t_2,t_f]-\Delta C_1^S[t_2,t_f]$ reads
\begin{equation}
    Q_1^S[t_2,t_f]~=~\frac{1}{\mu_1}\ln(1+e^{-\mu_1\eta})+\eta\frac{e^{-\mu_1\eta}}{1+e^{-\mu_1\eta}}.
\end{equation}
Since the system stays in energy eigenstates $\{|0\rangle,|1\rangle\}$ with vanishing eigenvalues, we have $\Delta E_S[t_2,t_f]=0$ during the time interval $[t_2,t_f]$.

We then successfully implement a quasi-static information erasure process that transforms the system from an initial mixed state $\rho_S(0)$ to a final pure state $\rho_S(t_f)$, with an entropy change $\Delta S(t_f)=S(t_f)-S(0)=-\ln 2$. Combining all steps (i)-(iv), we finally find
\bea
    &&\sum_{i=0}\lambda_{i}(0)\Delta C_i(t_f)+\ln(Z_{r}(t_f)/Z_{r}(0) )\nonumber\\
    &&=\sum_{i=0}\lambda_{i}(0)\Delta C_i(t_f)-\beta_BW_0^S[0,t_1]-\mu_1W_1^S[t_2,t_f]\nonumber\\
    &&=-\beta_B Q_0^S[0,t_1]-\mu_1 Q_1^S[t_2,t_f]+\beta_B\Delta E_S[t_1,t_2]\nonumber\\
    &&~~~+\mu_1\Delta C_1^S[t_1,t_2]\nonumber\\
    &&=-\ln2~=~\Delta S(t_f).
\eea
Here, we point out that $\Delta C_i(t_f)\equiv\Delta C_i[0,t_1]+\Delta C_i[t_1,t_2]+\Delta C_i[t_2,t_f]$.
This confirms that Eq. (\ref{eq:Our_ex}) indeed holds as an equality in the quasi-static limit, and also demonstrates that Eqs. (\ref{eq:Our_ex}) and (\ref{eq:LB_NJP_2}) coincide in this limit when a thermal bath in a generalized Gibbsian state is present.

\subsection{Numerical demonstration}
%===============================================
\begin{figure}[b!]
 \centering
\includegraphics[width=1\columnwidth]{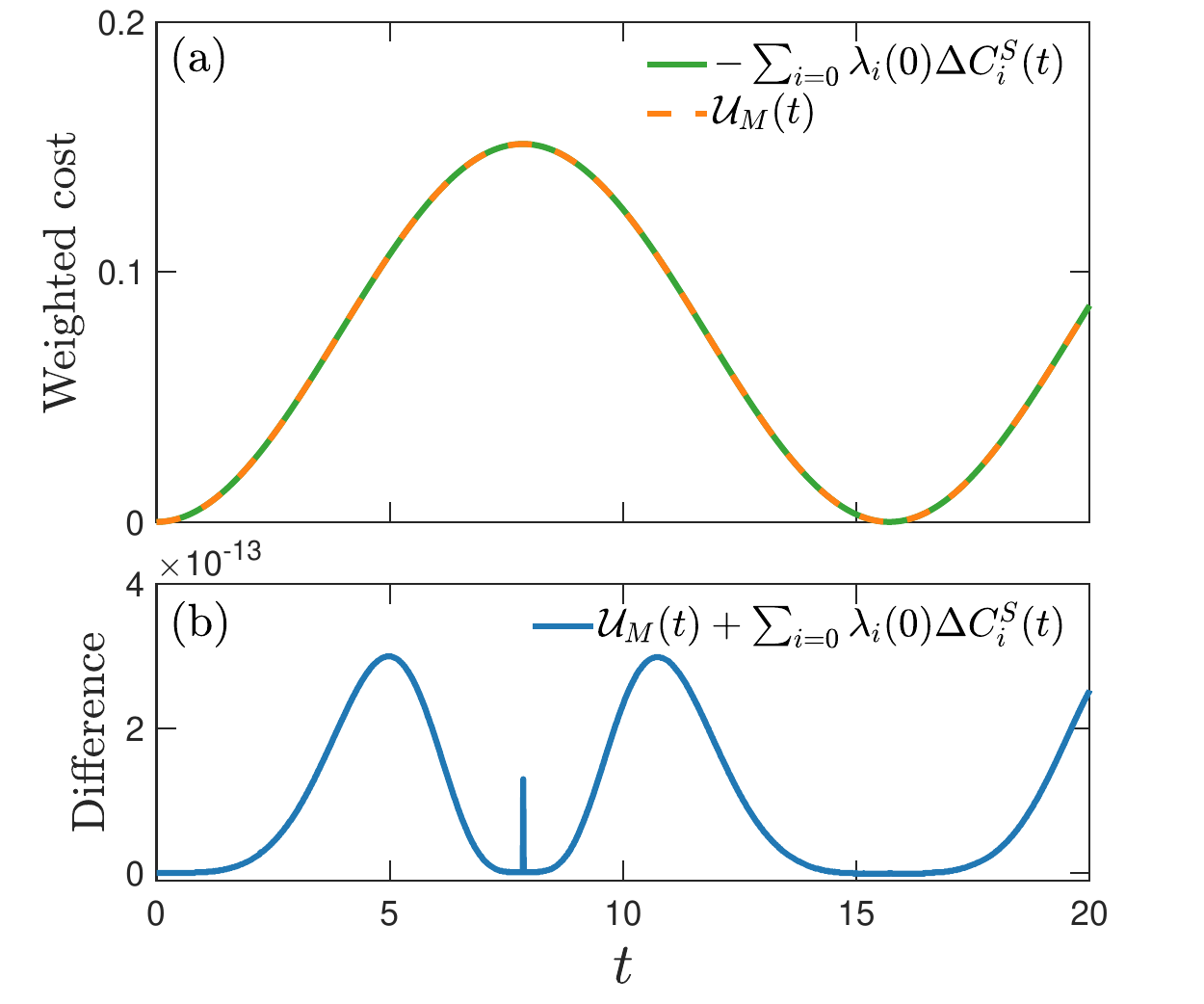} 
 \caption{(a) Time-dependent results for the weighted cost $-\sum_{i=0}\lambda_i(0)\Delta C_i^S(t)$ (green solid line) and the inferred upper bound $\mathcal{U}_M(t)$ (orange dashed line) given in Eq. (\ref{eq:Our_ex}). (b) Deviation  $\mathcal{U}_M(t)+\sum_{i=0}\lambda_i(0)\Delta C_i^S(t)$ as a function of time. Parameters are $\beta_A=0.5$, $\beta_B=2$, $\mu_A=0.5$, $\mu_B=1$, $\varepsilon=2$, $\eta=0.2$.}
\protect\label{fig:spin}
\end{figure}
%===============================================
To demonstrate the utility of Eq. (\ref{eq:Our_ex}) for finite-time quantum processes, we consider a minimum model of two conserved charges~\cite{Guan.25.PRA}. The model consists of two qubits described by the total Hamiltonian
\begin{equation}
    H~=~H_0+H_{\rm I}.
\end{equation}
Here, $H_0$ denotes the bare Hamiltonian for two qubits,
\begin{equation}
H_0~=~\sum_{j=A,B}H_j~=~-\sum_{j=A,B}\frac{\varepsilon_j}{2}\,\sigma_j^z.
\end{equation}
Here, $\varepsilon_j$ marks the energy gap of spin $j$ with $\sigma_j^z$ its Pauli matrix. We set $\varepsilon_A=\varepsilon_B=\varepsilon$ for simplicity. $H_{\rm I}$ represents a charge-conserving interaction between two spins
\begin{equation}
H_{\rm I}~=~ -\frac{\eta}{2}\bigl(\sigma_A^x\,\sigma_B^x+\sigma_A^y\,\sigma_B^y\bigr),
\end{equation}
where $\eta$ represents the spin–exchange strength and $\sigma_j^{x,y}$ denotes the corresponding Pauli matrices. It is straightforward to verify that the free and interaction Hamiltonian commute, $[H_0,H_{\rm I}]=0$, which implies $[\mathcal{N}_A+\mathcal{N}_B,H_{\rm I}]=0$, where $\mathcal{N}_j=(\mathbb{I}-\sigma_j^z)/2$ with $\mathbb{I}$ denoting a $2\times 2$ identity matrix is the excitation number operator of spin-$j$ \cite{Guimaraes.16.PRE,Lu.24.PRA,Guan.25.PRA}. These two commutation relations ensure that both the total mean energy and the total mean number of spin excitations are conserved during evolution.

The time evolution of the total density matrix $\rho(t)$ is governed by the Liouville–von Neumann equation $\partial\rho(t)/\partial t=-i[H_0+H_{\rm I},\rho(t)]$. We consider an initial product state for the total system, $\rho(0)=\gamma_A\otimes\gamma_B$, with each spin prepared in a local grand-canonical state ($j=A,B$),
\begin{equation}
  \gamma_j~=~\frac{e^{-\beta_j\,(H_j - \mu_j \mathcal{N}_j)}}{\mathcal{Z}_i}.  
\end{equation}
Here, $\beta_j$ and $\mu_j$ represent, respectively, the initial inverse temperature and the effective potential conjugate to the excitation number of spin $j$. In this spin setup, $\mu_j$ plays the role of an adjustable parameter controlling the initial population of excitations. $\mathcal{Z}_j = \text{Tr}\left[\exp\left(-\beta_j (H_j - \mu_j \mathcal{N}_j)\right)\right]$ is the partition function corresponding to each spin. When $\beta_A\neq\beta_B$ and $\mu_A\neq\mu_B$, we expect exchange of charges between two spins. 

To demonstrate Eq. (\ref{eq:Our_ex}), we treat spin $A$ as the system and the remaining spin $B$ as its finite-sized environment, such that $C_0^S(t)=E_A(t)=\mathrm{Tr}[\rho_A(t)H_A]$ and $C_1^S(t)=N_A(t)=\mathrm{Tr}[\rho_A(t)\mathcal{N}_A]$. We remark that this model already lies beyond the validity regime of the generalized Landauer lower bound in Eq. (\ref{eq:LB_NJP}), as the finite-sized environment cannot sustain the initial generalized Gibbsian state throughout the process. 

We track the evolution of spin $A$ to verify Eq. (\ref{eq:Our_ex}). The form of initial local state implies that the initial Lagrange multipliers are $\lambda_0(0)=\beta_A$ and $\lambda_1(0)=-\beta_A\mu_A$ in Eq. (\ref{eq:reference_state}). o determine time-dependent Lagrange multipliers needed to evaluate $\mathcal{R}(t)$ in Eq. (\ref{eq:Our_ex}), we proceed as follows: We numerically solve the Liouville–von Neumann equation to obtain the reduced state $\rho_A(t)$ from the full state $\rho(t)$. Using $\rho_A(t)$, we compute the expectation values $E_A(t)$ and $N_A(t)$, from which $\lambda_{0,1}(t)$ are obtained by solving the coupled equations $\mathrm{Tr}[\rho_r^A(t)H_A]=E_A(t)$ and $\mathrm{Tr}[\rho_r^A(t)\mathcal{N}_A]=N_A(t)$ with $\rho_r^A(t)=\exp[-\lambda_0(t)H_A - \lambda_1(t)\mathcal{N}_A]/\mathcal{Z}_r^A(t)$. We depict a set of results in Fig. \ref{fig:spin} that confirms the validity of Eq. (\ref{eq:Our_ex}) in this minimum spin model. Notably, the deviation between the upper bound $\mathcal{U}_M(t)$ and the weighted sum $-\sum_{i=0}\lambda_i(0)\Delta C_i^S(t)$ is negligible in this minimum model, implying that the system remains very close to a grand-canonical state throughout the evolution under the chosen parameter set.

%=========================================================
\section{Quantum systems knowing fluctuations}\label{sec:3}
Previously, we obtained and examined an information-cost relation for quantum systems in which only the mean values of charges are accessible. We adopted the conventional perspective--prevalent in existing literature--that interprets changes in the mean values of these charges as the relevant thermodynamic cost for completing quantum processes. This convention originates largely from the LP which was initially formulated in the context of classical information erasure process where mean values are indeed the quantities of primary interest. However, at the nanoscale, thermodynamic fluctuations of observables become ubiquitous and significant--yet they lie beyond the scope of current formulations of information-cost relations.

In order to develop a more complete description of quantum processes, it is essential to account for changes not only in the mean values but also in the higher-order fluctuations of observables. To illustrate this point, consider the following example: A single qubit with Hamiltonian $H_S=\omega \sigma_z/2$ couples to a zero-temperature bath and undergoes an {\it ideal} qubit reset process. The system starts from the maximally mixed state $\rho_i=\frac{1}{2}\mathrm{I}$ (where $\mathrm{I}$ denotes a $2\times 2$ identity matrix) and ends in the ground state $\rho_f=|g\rangle\langle g|$, with $\sigma_z|g\rangle=-|g\rangle$. The energy change during the process is $E_i-E_f=\omega/2$, where $E_{i,f}\equiv\mathrm{Tr}[\rho_{i,f}H_S]$. This energy loss is dissipated as heat and satisfies the Landauer lower bound \cite{Timpanaro.20.PRL}. However, attention must also be paid to the change in energy fluctuation. One first notes that the final pure state renders a vanishing energy fluctuation $\mathrm{Var}[H_S]_f\equiv\mathrm{Tr}[\rho_f(H_S-E_f\cdot \mathrm{I})^2]=0$, while the initial energy fluctuation is $\mathrm{Var}[H_S]_i=\omega^2/4$. Hence, we also expect a finite change in energy fluctuation $\mathrm{Var}[H_S]_i-\mathrm{Var}[H_S]_f=\omega^2/4$ which can be comparable in magnitude to the energy change itself. Therefore, to successfully drive the system through such a designed quantum process, it is not sufficient to compensate only the first-order energy changes; adjustment of higher-order fluctuations are also required to ensure accurate realization of the target quantum dynamics. For later convenience, we propose to interpret the change in fluctuation as a fluctuation cost related to the quantum process, thereby extending the conventional notion of thermodynamic cost beyond mean values.

By analogy with the LP, which bounds the first-order thermodynamic cost using the information content change $\Delta S$, it is natural to ask whether analogous bounds constrain the thermodynamic fluctuation cost in relation to $\Delta S$. However, conventional thermodynamic frameworks--primarily built on entropic inequalities such as the second law of thermodynamics--currently lack the capacity to reveal such fluctuation cost bounds.

%===========================================================
\subsection{Information-fluctuation cost relation}
In the following, we assume that the second-order fluctuations (variances) of charges are also experimentally accessible. A close examination of the thermodynamic inference strategy introduced in Sec. \ref{sec:1} shows that the dynamical constraints used to derive the maximum-entropy state in Eq. (\ref{eq:reference_state}) can included any observations which naturally encompass measurable variances (see, e.g., Ref.~\cite{Herrera.21.PRL}). Therefore, we can directly apply the thermodynamic inference framework to derive information-thermodynamic cost relations that can establish bounds on fluctuation costs. To make the derivation as clear as possible, we explicitly consider a single charge $\mathcal{C}_i^S$ whose mean value and variance are known in what follows. Generalization to higher-order fluctuations is conceptually straightforward, though practically constrained by experimental capabilities.

For an arbitrary charge $\mathcal{C}_i^S$ with a time-dependent mean value $\langle\mathcal{C}_i^S\rangle(t)$, its time-dependent variance is defined as:
\begin{equation}
\mathrm{Var}[\mathcal{C}_i^S](t)~\equiv~\mathrm{Tr}\left[\rho_S(t)\left(\mathcal{C}_i^S- \langle\mathcal{C}_i^S\rangle(t)\mathrm{I}\right)^2\right].
\end{equation}
Here, $\mathrm{I}$ is the identity matrix acting on the same space as $\mathcal{C}_i^S$. From this definition, it is clear that the operator associated with the variance reads $\mathcal{V}_i^S(t)\equiv\left(\mathcal{C}_i^S- \langle\mathcal{C}_i^S\rangle(t)\mathrm{I}\right)^2$. Substituting this operator into Eq. (\ref{eq:reference_state}), we receive a reference state with a quadratic exponent--dubbed quadratic reference state hereafter--based on the maximum entropy principle (see also a detailed derivation based on the method of Lagrange multipliers in Appendix~\ref{a:1}),
\begin{equation}\label{eq:gaussian}
\rho_r(t)~=~\frac{\exp\left[-\lambda_m(t)\mathcal{C}_i^S-\lambda_v(t) \mathcal{V}_i^S(t)\right]}{Z_r(t)},
\end{equation}
where $Z_r(t) =\mathrm{Tr}\left\{\exp\left(-\lambda_m(t)\mathcal{C}_i^S-\lambda_v(t) \mathcal{V}_i^S(t)\right)\right\}$ is the corresponding normalization factor. The Lagrange multipliers $\lambda_m(t)$ and $\lambda_v(t)$ are determined by requiring that the mean and variance computed with the reference state match the actual mean and variance, $\mathrm{Tr}\left[\rho_r(t)\mathcal{C}_i^S\right]=\langle\mathcal{C}_i^S\rangle(t)$ and $\mathrm{Tr}\left[\rho_r(t)\mathcal{V}_i^S(t)\right]=\mathrm{Var}[\mathcal{C}_i^S](t)$.
%We remark that this Gaussian reference state incorporates information about both the mean and the variance of the observable $\mathcal{C}_i^S$, with only the Lagrange multiplier associated with the variance playing a role (see Appendix~\ref{a:1}).

Since the quadratic reference state in Eq. (\ref{eq:gaussian}) takes an exponential form just as the Gibbsian reference state presented earlier, we can easily check that Eqs. (\ref{eq:s_devi}) and (\ref{eq:equality}) remain applicable by just replacing arbitrary observables $\mathcal{O}_i$ with $\mathcal{C}_i^S$ and its variance-associated operator $\mathcal{V}_i^S(t)$. Hence, we can directly generalize Eq. (\ref{eq:Our_ex}) to obtain a thermodynamic inequality, 
\begin{equation}\label{eq:fluctuation_inequality}
\lambda_{v}(0)\Delta \mathrm{Var}[\mathcal{C}_i^S](t)+\mathcal{R}_v(t)~\ge~\Delta S(t).
\end{equation}
Here, we have denoted $\mathcal{R}_v(t)\equiv\Delta\lambda_{v}(t)\cdot\mathrm{Var}[\mathcal{C}_i^S](t)+\Delta S^{in}+\ln(Z_{r}(t)  /Z_{r}(0))+\lambda_m(t)C_i^S(t)-\lambda_m(0)C_i^S(0)$
where $\Delta S^{in}$ denotes the contrast in von Neumann entropy between the initial state $\rho_S(0)$ and its corresponding quadratic reference state $\rho_r(0)$, $\Delta\lambda_{v}(t)=\lambda_{v}(t)-\lambda_{v}(0)$, $Z_r(t)$ is defined in Eq. (\ref{eq:gaussian}). Following the derivation, we know that saturation of Eq. (\ref{eq:fluctuation_inequality}) requires $D[\rho_S(t)||\rho_r(t)]=0$. In this scenario, the reference state $\rho_r(t)$ in Eq. (\ref{eq:gaussian}) has a quadratic exponent. This implies that the saturation condition of Eq. (\ref{eq:fluctuation_inequality}) differs from that of Eq. (\ref{eq:Our_ex}) since the thermalized state in the asymptotic limit does not necessarily have a quadratic exponent. Rather, saturation of Eq. (\ref{eq:fluctuation_inequality}) requires vanishing higher-order fluctuations beyond the variance--a condition naturally satisfied by Gaussian continuous-variable systems~\cite{Genoni.16.CP,Braunstein.05.RMP,Weedbrook.12.RMP}. For general non-Gaussian systems, we expect strict saturation to be unattainable, as higher-order fluctuations are ubiquitous at the nanoscale.

We emphasize that Eq. (\ref{eq:fluctuation_inequality}) provides a desired information-cost relation relating the fluctuation cost $\Delta \mathrm{Var}[\mathcal{C}_i^S](t)$ and the system's information content change $\Delta S(t)$. Eq. (\ref{eq:fluctuation_inequality}) thus represents our second main result on information-cost relations. To facilitate numerical comparison, we rearrange Eq. (\ref{eq:fluctuation_inequality}) to obtain a lower bound on the fluctuation cost $\Delta \mathrm{Var}[\mathcal{\mathcal{C}_i^S}](t)$, 
\begin{equation}\label{eq:var_eq}
\Delta \mathrm{Var}[\mathcal{C}_i^S](t)~\ge~[\Delta S(t)-\mathcal{R}_v(t)]\lambda_{v}^{-1}(0)~\equiv~\mathcal{L}_v(t).
\end{equation}
To our knowledge, the connection between fluctuation change and system information content change expressed in Eq. (\ref{eq:fluctuation_inequality}) [or Eq. (\ref{eq:var_eq})] has not been appreciated before. We note that the celebrated thermodynamic uncertainty relation (TUR)~\cite{Barato.15.PRL,Gingrich.16.PRL,Horowitz.19.NP,Liu.19.PRE} also establishes a connection between variance and entropy. However, when considering which formulation more naturally extends the spirit of the LP to second-order fluctuations, several key distinctions between the TUR and Eq. (\ref{eq:fluctuation_inequality}) should be emphasized: (i) The TUR addresses instantaneous relative fluctuations weighted by the mean values. This contrasts in spirit with the LP which addresses energy change in the form of heat dissipation. In contrast, Eq. (\ref{eq:fluctuation_inequality}) considers the change in absolute fluctuations over a time interval, aligning more closely with the LP. (ii) The TUR involves the total entropy production, requiring the knowledge of the environmental information to evaluate, while the LP concerns only the system entropy change. In comparison, Eq. (\ref{eq:fluctuation_inequality}) relies on the system entropy change and remains applicable even in the presence of unknown or non-thermal environments. Hence, we argue that Eq. (\ref{eq:fluctuation_inequality}) more appropriately generalizes the spirit of the LP to second-order fluctuations.

\subsection{Numerical demonstration I: Single qubit erasure process}
To numerically verify Eq. (\ref{eq:fluctuation_inequality}), we first consider a driven qubit immersed in a single thermal bath with temperature $T_E$ which was used to realize an information erasure process \cite{Miller.20.PRL,Saito.22.PRL,Liu.23.PRAa,LiuJ.24.PRR}. The driven qubit is described by a time-dependent Hamiltonian
\begin{equation}\label{eq:hst}
H_S(t)~=~\frac{\varepsilon(t)}{2}\Big(\cos[\theta(t)]\sigma_z+\sin[\theta(t)]\sigma_x\Big).
\end{equation}
Here, $\sigma_{x,z}$ denote the Pauli matrices, and we have two driving fields $\varepsilon(t)$ and $\theta(t)$. We adopt the driving protocols $\varepsilon(t)=\varepsilon_0+(\varepsilon_{\tau}-\varepsilon_0)\sin(\pi t/2\tau)^2$ and $\theta(t)=\pi(t/\tau-1)$ \cite{Miller.20.PRL} with $\tau$ being the time duration of the information erasure process. Setting $\varepsilon_{\tau}\gg \varepsilon_0$ and $\varepsilon_{\tau}/T_E\gg 1$, one can reset the qubit to a final state that is very close to its ground state, thus realizing an information erasure process.

%==========================================
\begin{figure}[b!]
 \centering
\includegraphics[width=1\columnwidth]{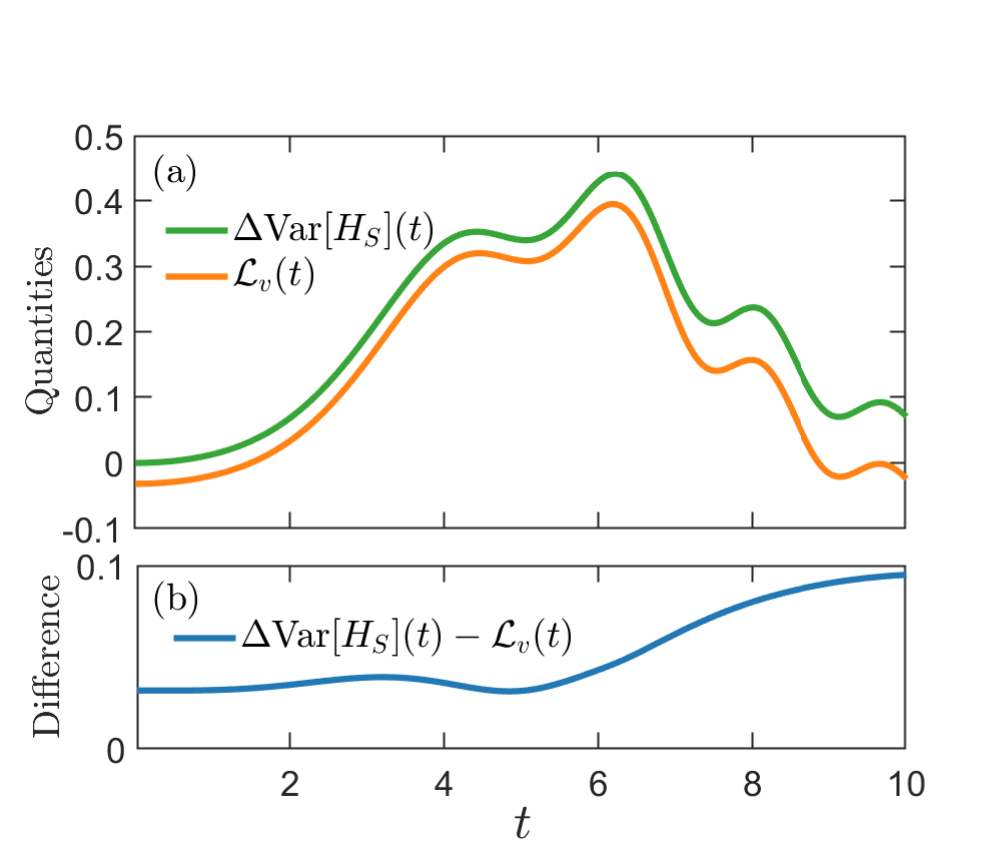} 
 \caption{(a) Results for the energy fluctuation cost $\Delta \mathrm{Var}[H_S](t)$ (green solid line) and the inferred lower bound $\mathcal{L}_v(t)$ (orange solid line) [cf. Eq. (\ref{eq:var_eq})] as functions of time. (b) Deviation of $\Delta \mathrm{Var}[H_S](t)-\mathcal{L}_v(t)$. Parameters are $\varepsilon_0=0.4$, $\varepsilon_{\tau}=4$, $\tau=10$, $\gamma_{1,2}=0.2$, and $T_E=0.25$.}
\protect\label{fig:AET}
\end{figure}
%==========================================
The evolution of the actual system state $\rho_S(t)$ is governed by a quantum Lindblad master equation \cite{Saito.22.PRL}
\begin{equation}\label{eq:lindblad}
 \frac{d}{dt}\rho_S(t)~=~-i[H_S(t),\rho_S(t)]+\sum_{\mu=1}^2\gamma_{\mu}\mathcal{D}[L_{\mu}(t)]\rho_S(t). 
 \end{equation}
Here, $\gamma_{\mu}\geqslant 0$ is the damping coefficient of decaying channel $\mu$, $\mathcal{D}[L_{\mu}]\rho=L_{\mu}\rho L_{\mu}^{\dagger}-\frac{1}{2}\{L_{\mu}^{\dagger}L_{\mu},\rho\}$ denotes a Lindblad superoperator with $\{\mathcal{O}_1,\mathcal{O}_2\}=\mathcal{O}_1\mathcal{O}_2+\mathcal{O}_2\mathcal{O}_1$. We have two time-dependent jump operators $L_1(t)=\sqrt{\varepsilon(t)[N_E(t)+1]}|0_t\rangle\langle 1_t|$, $L_2(t)=\sqrt{\varepsilon(t)N_E(t)}|1_t\rangle \langle 0_t|$ describing de-excitation and excitation process induced by the thermal bath, respectively. Here $|0_t\rangle$ ($|1_t\rangle$) is the instantaneous ground (excited) state of $H_S(t)$, $N_E(t)=1/[e^{\varepsilon(t)/T_E}-1]$ is the Bose-Einstein distribution.
% the heat dissipation of the setup has been analyzed using the simplified upper bound previously.

For a demonstration purpose, we consider the system energy and its variance to be observations with $\mathcal{C}_0^S=H_S$ in Eq. (\ref{eq:var_eq}). The Lagrange multipliers $\lambda_m(t)$ and $\lambda_v(t)$ are determined by numerically solving the equations $\mathrm{Tr}[\rho_r(t)H_S(t)]=E_S(t)$ and $\mathrm{Tr}[\rho_r(t)(H_S(t)-E_S(t))^2]=\mathrm{Var}[H_S(t)]$, where $E_S(t)$ and $\mathrm{Var}[H_S(t)]$ are obtained by evolving Eq. (\ref{eq:lindblad}). A set of numerical results for Eq. (\ref{eq:var_eq}) with an initial system Gibbsian state $\rho_S(0)=e^{-H_S(0)/T_E}/\mathrm{Tr}[e^{-H_S(0)/T_E}]$ is presented in Fig. \ref{fig:AET}. From Fig. \ref{fig:AET} (a) and (b), one can observe that the energy fluctuation change $\Delta \mathrm{Var}[H_S](t)$ is consistently lower-bounded by the corresponding $\mathcal{L}_v(t)$, thereby demonstrating the validity of Eq. (\ref{eq:var_eq}) and equivalently, Eq. (\ref{eq:fluctuation_inequality}). Moreover, as shown in Fig. \ref{fig:AET} (b), the difference between $\Delta \mathrm{Var}[H_S](t)$ and $\mathcal{L}_v(t)$ is relatively small compared to their magnitudes in Fig. \ref{fig:AET} (a), indicating that the quadratic reference state derived from the maximum entropy principle effectively captures the finite-time behavior of the actual energy fluctuations. Notably, the discrepancy between $\Delta \mathrm{Var}[H_S](t)$ and $\mathcal{L}_v(t)$ gradually diminishes over time, implying that the system progressively converges toward the quadratic reference state under the evolution and higher-order fluctuations of energy become negligible at long times.

% For slow but finite-time processes, the fluctuation is Gaussian (Jarzynski.97.PRL), we expect reducing difference as the time duration $\tau$ of process increases. 

\subsection{Numerical demonstration II: Driven inelastic heat transfer process}
We emphasize that Eqs. (\ref{eq:fluctuation_inequality}) and (\ref{eq:var_eq}) are valid for arbitrary quantum nonequilibrium processes, extending beyond the information erasure process to which the LP is conventionally applied. To illustrate this generality, we consider a driven three-terminal quantum dot system~\cite{JiangCRP,JiangReview}. The total setup consists of a double quantum dot system (specified as the $L$ and $R$ dots) coupled to a phononic thermal bath, while each dot further individually exchanges electrons with an electronic thermal reservoir. The total Hamiltonian for the setup contains three different parts. The first part of the total Hamiltonian describes the ``bare" system--two coupled quantum dots whose site energies can be tuned,
\begin{equation}
H_{\rm DQD}(t) = \sum_{i=L,R} \varepsilon_i(t)  d_i^\dagger  d_i +  {\delta}( d_L^\dagger  d_R + {\rm H.c.}).
\end{equation}
Here, `$\mathrm{H.c.}$' denotes the Hermitian conjugate of the term in the bracket, $d_i$ and $\varepsilon_i(t)$ denote an electronic annihilation operator and a time-dependent site energy of the $i$-th quantum dot, respectively. We adopt the driving protocols $\varepsilon_L(t) = \varepsilon_{L0}+\varepsilon_{L\tau}\sin(\Omega t)$, $\varepsilon_R(t) = \varepsilon_{R0}+ \varepsilon_{R\tau}\sin(\Omega t + \phi)$ with $\Omega$ and $\phi$ being the frequency and the nonzero modulation phase, respectively~\cite{LuPRB24}. This simple set of driving protocols is sufficient for a demonstration purpose. $\delta$ measures the tunneling strength between the two dots. The second part of the total Hamiltonian accounts for a phononic thermal bath containing an ensemble of harmonic oscillators, and their inelastic coupling to the two dots. The $q$-th harmonic oscillator has frequency $\omega_q$ with annihilation operator $a_q$ and $q$-dependent electron-phonon coupling strength $\lambda_q$, resulting in a Hamiltonian denoting the second part $H_{ep}=\sum_q\left[\omega_{q}a^\dagger_q a_q/2+\lambda_q d_L^\dagger d_R (a_q + a^\dagger_q) + {\rm H.c.}\right]$~\cite{Jiang2012,MyPRBMultitask}. The third part of the total Hamiltonian describes two electronic reservoirs with electronic annihilation operators $\{d_{vk}\}$ and energies $\{\varepsilon_{vk}\}$ as well as their coupling to the individual dot measured by a mode-dependent coupling strength $\{\gamma_{vk}\}$, leading to a Hamiltonian describing the third part $H_{e-\mathrm{lead}}=\sum_{v=L,R}\sum_{k} \left(\varepsilon_{vk} d_{vk}^\dagger d_{vk}/2+\gamma_{vk} d_v^\dagger d_{vk} +  {\rm H.c.}\right)$. The influence of phononic and electronic baths is captured by the spectral functions (or hybridization energies) $\Gamma_{\rm ph}=2\pi \sum_q {\lambda^2_q} \delta(\omega-\omega_q)$ and $\Gamma_{v}=2\pi\sum_k |\gamma_{vk}|^2 \delta(\varepsilon-\varepsilon_{vk})$ ($v=L,R$), respectively. For simplicity, they are chosen to be flat (wide-band limit).

%
%-------------------------------------------------
\begin{figure}[b!]
\centering
\includegraphics[width=1\columnwidth]{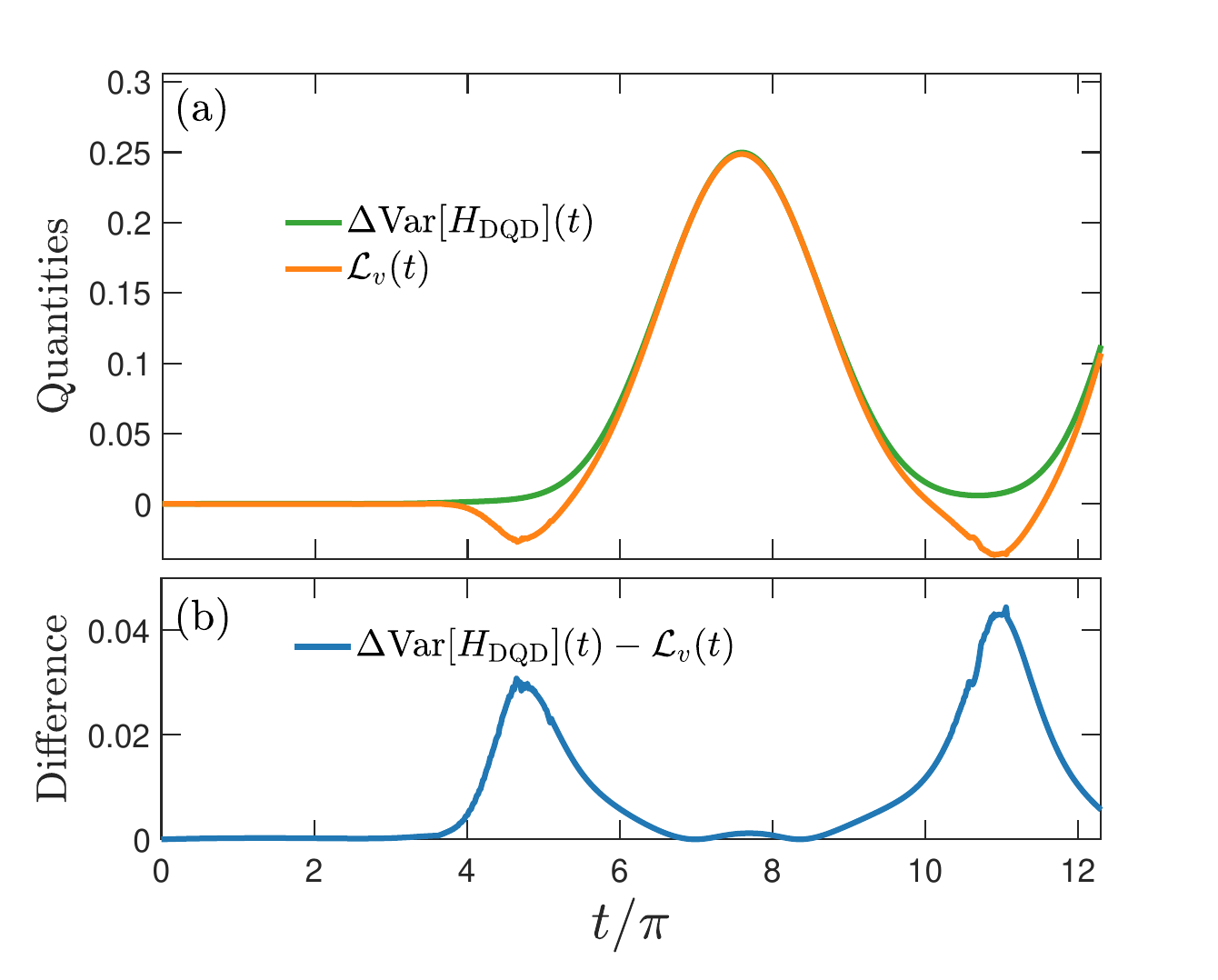}
\caption{(a) Results for the energy fluctuation cost $\Delta \mathrm{Var}[H_{\rm{DQD}}](t)$ (green solid line) and the inferred lower bound $\mathcal{L}_v(t)$ (orange solid line) [cf. Eq. (\ref{eq:var_eq})] as functions of time. (b) Deviation of $\Delta \mathrm{Var}[H_{\rm{DQD}}](t)-\mathcal{L}_v(t)$. Parameters are $\mu_L=\mu_R=0$, $\Omega=1$, $\varepsilon_{L0}=2$, $\varepsilon_{R0}=2$, $\varepsilon_{L\tau}=1.5$, $\varepsilon_{R\tau}=1.5$, $\phi=\pi/10$, $\Gamma_L=\Gamma_R=\Gamma_{\rm ph}=0.1$, $\delta=0.2$, $T_L=0.2$, $T_R=0.3$, $T_{\rm ph}=0.4$ and the initial system Gibbsian state is fixed by setting $T_0=0.125.$}
\label{fig:DQD}
\end{figure}
%-------------------------------------------------
We consider the weak coupling limit in which both the electron-phonon and dot-reservoir couplings are treated up to second order. Introducing the interaction Hamiltonian $V=\sum_q\lambda_q d_L^\dagger d_R (a_q + a^\dagger_q)+\sum_{v=L,R}\sum_{k}\gamma_{vk} d_v^\dagger d_{vk}+{\rm H.c.}$, the dynamics of the reduced system state $\rho_S$ is described by the quantum Redfield master equation~\cite{Breuer.02.NULL}
\bea\label{eq:redfield}
\frac{\partial}{\partial t} {\rho}_S(t)
&=& i[{\rho}_S(t), H_{\rm DQD}(t)]\nonumber\\
&&-\int_0^\infty d\tau {\rm Tr}_E\Big\{[[V,[V(-\tau),
{\rho}_S(t){\rho}_E]
]\Big\}.                         
\eea
Here, ${\rho}_S(t){\rho}_E$ is short for ${\rho}_S(t)\otimes{\rho}_E$, $\rho_E=\rho_L{\otimes}\rho_R{\otimes}\rho_{\rm ph}$ denotes an initial product thermal state for the phononic and electronic baths
where $\rho_{\rm ph}=\exp\left[-\beta_{\rm ph} H_{\rm ph}\right] / \text{Tr}\left\{\exp\left[-\beta_{\rm ph} H_{\rm ph}\right]\right\}$ and
$\rho_v=\exp\left[-\beta_{v}(H_{v}-\mu_v N_v)\right]/\text{Tr}\left\{\exp\left[-\beta_{v}(H_{v} - \mu_v N_v)\right]\right\}~(v=L,R)$ with
$H_{\rm ph}=\sum_q\omega_{q}a^\dagger_qa_q$, $H_v=\sum_k\varepsilon_{vk} d_{vk}^\dagger d_{vk}$ and $N_v=\sum_kd_{vk}^{\dagger}d_{vk}$. $\beta_{\rm ph}=T_{\rm ph}^{-1}$ and $\beta_v=T_v^{-1}$ denote inverse bath temperatures. $\mu_v$ are chemical potentials of the electronic reservoirs. 

We note that the setup can be configured to operate as an inelastic heat engine \cite{LuPRB24}. Here, however, we focus solely on demonstrating the validity of Eq. (\ref{eq:fluctuation_inequality}) or (\ref{eq:var_eq}) in a general scenario without engineering control fields. We still consider system energy and its variance to be the observables with $\mathcal{C}_0^S=H_{\rm{DQD}}$ in Eq. (\ref{eq:var_eq}). The Lagrange multipliers $\lambda_m(t)$ and $\lambda_v(t)$ are determined by numerically solving the equations $\mathrm{Tr}[\rho_r(t)H_{\rm{DQD}}(t)]=E_{\rm{DQD}}(t)$ and $\mathrm{Tr}[\rho_r(t)(H_{\rm{DQD}}(t)-E_{\rm{DQD}}(t))^2]=\mathrm{Var}[H_{\rm{DQD}}(t)]$, where $E_{\rm{DQD}}(t)$ and $\mathrm{Var}[H_{\rm{DQD}}(t)]$ are obtained by evolving Eq. (\ref{eq:redfield}). A set of numerical results for the energy fluctuation change $\Delta \mathrm{Var}[H_{\rm{DQD}}](t)$ and its lower bound $\mathcal{L}_v(t)$ from Eq. (\ref{eq:var_eq}) is depicted in Fig. \ref{fig:DQD}, using an initial system Gibbsian state $\rho_S(0)=e^{-H_{\rm{DQD}}(0)/T_0}/\mathrm{Tr}[e^{-H_{\rm{DQD}}(0)/T_0}]$. In Fig. \ref{fig:DQD} (a), we depict numerical results for the energy fluctuation change $\Delta \mathrm{Var}[H_{\rm{DQD}}](t)$ for the driven quantum dot model during the process, along with the predicted lower bound $\mathcal{L}_v(t)$ given in Eq. (\ref{eq:var_eq}). Together with results for the difference $\Delta \mathrm{Var}[H_{\rm{DQD}}](t)-\mathcal{L}_v(t)$ showed in Fig. \ref{fig:DQD} (b), we confirm the validity of Eq. (\ref{eq:var_eq}) in this model.

\section{Discussion and conclusion}\label{sec:4}
In summary, we have developed a thermodynamic inference approach based on the maximum entropy principle~\cite{Jaynes.57.PR} to analyze thermodynamic cost–information trade-off relations in arbitrary nonequilibrium quantum systems. Our framework extends the scope of the Landauer’s principle and its subsequent generalizations in both the nature of the derived relations and the range of applicability.

Regarding the revealed relations, we present two main results that highlight the distinct advantages of our inference approach. First, for quantum systems in which only mean values of charges are accessible, we derived a general thermodynamic upper bound [Eq. (\ref{eq:Our_ex})] on the thermodynamic cost in forms of system's charge loss, which complements an existing generalized Landauer lower bound~\cite{Lostaglio.17.NJP} in its regime of validity. We demonstrated that the upper and lower bounds converge in the quasi-static limit, consistent with the saturation of the second law of thermodynamics. Second, for scenarios where higher-order fluctuations such as variances of charges become further measurable, we introduced the concept of fluctuation cost and derived an entropy-informed thermodynamic lower bound for it [Eq. (\ref{eq:var_eq})]. This result captures genuinely quantum nature of nanoscale processes where fluctuations are significant, going beyond conventional Landauer-type bounds that address only first-order means. Both results were numerically verified using distinct models undergoing different quantum processes.

Concerning the application range, our approach is independent of environmental details, relying solely on available system-level information for thermodynamic inference. This contrasts with traditional Landauer-type bounds, which typically require a thermal bath assumption. Our results are therefore applicable to a broader class of nonequilibrium quantum systems--including those coupled to non-thermal or even unknown environments--situations where existing bounds fail. This makes our framework particularly relevant for quantum technological applications where environments may be engineered~\cite{Harrington.22.NRP} or are not well characterized.

Looking forward, potential extensions of this work include incorporating non-Markovian effects, analyzing systems with non-commuting charges, and experimental validation using established platforms such as superconducting qubits.

\section*{Acknowledgments}
We thank the anonymous reviewers for their valuable comments and suggestions that helped improve this work. J. L. acknowledges support from the National Natural Science Foundation of China (Grant No. 12205179), the Shanghai Pujiang Program (Grant No. 22PJ1403900) and start-up funding of Shanghai University. J.-H. Jiang acknowledges support from the National Natural Science Foundation of China (Grant No. 12125504) and the ``Hundred Talents Program'' of the Chinese Academy of Sciences.

\appendix
%=============================================================
\renewcommand{\theequation}{A\arabic{equation}}
\renewcommand{\thefigure}{A\arabic{figure}}
\setcounter{equation}{0}  % reset counter
\setcounter{figure}{0}  % reset counter
\section{Derivation of quadratic reference state via maximum entropy principle}
\label{a:1}
The maximum entropy principle provides a powerful method for inferring the least biased state consistent with a given set of constraints. In the following, we will show how to get the form of quadratic reference state in Eq. (\ref{eq:gaussian}) of the main text given the mean value and the variance of a charge $\mathcal{C}_i^S$ are known. For simplicity, we just denote the corresponding mean value as $\mu$, and the variance as $\sigma^2$. We also suppress possible time-dependence in this appendix.

Following Ref.~\cite{Jaynes.57.PR}, we still utilized the method of Lagrange multipliers in which we maximize the system's von Neumann entropy subject to the given constraints. Let $\rho$ be a system density matrix, its von Neumann entropy is defined as
\begin{equation}
S[\rho] = -\mathrm{Tr}[\rho \ln \rho].
\end{equation}
For the current scenario, we need to impose the following three constraints on the system:
\begin{enumerate}
    \item \textbf{Normalization:} The total probability must be unity
        \begin{equation}
        \mathrm{Tr}[\rho] = 1.
        \end{equation}
    \item \textbf{Fixed Mean:} The expectation value of the observable $\mathcal{C}_i^S$ is fixed to $\mu$ with respect to $\rho$
        \begin{equation}
        \mathrm{Tr}[\rho \mathcal{C}_i^S] = \mu.
        \end{equation}
    \item \textbf{Fixed Variance:} The variance $\mathcal{C}_i^S$ is fixed to $\sigma^2$
        \begin{equation}
        \mathrm{Tr}[\rho(\mathcal{C}_i^S-\mu\mathrm{I})^2] = \sigma^2.
        \end{equation}
        Here, $\mathrm{I}$ is the identity matrix.
\end{enumerate}

To apply the maximum entropy principle, we construct the Lagrangian functional $\mathcal{L}[\rho]$ by introducing Lagrange multipliers for each constraint, let $\lambda_1$ be the multiplier for normalization, $\lambda_2$ for the mean, and $\lambda_3$ for the variance,
\bea
 \mathcal{L}[\rho] &=& -\mathrm{Tr}[\rho \ln \rho] + \lambda_1 \left( \mathrm{Tr}[\rho] - 1 \right)+ \lambda_2 \left( \mathrm{Tr}[\rho \mathcal{C}_i^S] - \mu \right)\nonumber\\
 && + \lambda_3 \left( \mathrm{Tr}[\rho(\mathcal{C}_i^S-\mu\mathrm{I})^2] - \sigma^2\right).
\eea
To find the maximum, we take its functional derivative with respect to $\rho$ to zero
\begin{equation}
\frac{\delta \mathcal{L}}{\delta \rho} = 0.
\end{equation}

Noting that the variation of the entropy term is $\delta[-\mathrm{Tr}(\rho\ln\rho)]=-\mathrm{Tr}[\delta\rho\ln\rho+\delta\rho]$. Substituting this into the variation of the Lagrangian, we get 
\begin{equation}
    -\mathrm{Tr}\left(\delta\rho[\ln\rho+(1-\lambda_1)\mathrm{I}-\lambda_2\mathcal{C}_i^S-\lambda_3(\mathcal{C}_i^S-\mu\mathrm{I})^2]\right)=0.
\end{equation}
Since this must hold for any arbitrary variation $\delta\rho$, the term inside the square brackets must be zero,
\begin{equation}
\ln \rho +(1 -\lambda_1)\mathrm{I} - \lambda_2 \mathcal{C}_i^S - \lambda_3 (\mathcal{C}_i^S -\mu\mathrm{I})^2 = 0.
\end{equation}
We solve for $\rho$ by exponentiating both sides, yielding an inferred density matrix
\begin{equation}
\rho_r ~=~ \exp\left[ -(1 - \lambda_1)\mathrm{I} + \lambda_2 \mathcal{C}_i^S + \lambda_3 (\mathcal{C}_i^S-\mu\mathrm{I})^2 \right].
\end{equation}
We can separate the normalization constant (partition function $Z_r$) from the operator part. Let $Z_r=\exp[1-\lambda_1]$, we can rewrite the inferred state in a more compact form
\begin{equation}
\rho_r~=~\frac{1}{Z_r}\exp\left[ \lambda_2 \mathcal{C}_i^S + \lambda_3 (\mathcal{C}_i^S-\mu\mathrm{I})^2 \right],
\end{equation}
Which is just Eq. (\ref{eq:gaussian}) in the main text. Clearly, $Z_r=\mathrm{Tr}[\exp(\lambda_2 \mathcal{C}_i^S + \lambda_3 (\mathcal{C}_i^S-\mu\mathrm{I})^2)]$ since $\mathrm{Tr}[\rho_r]=1$. We also require $\mathrm{Tr}[\rho_r\mathcal{C}_i^S]=\mu$ and $\mathrm{Tr}[\rho_r(\mathcal{C}_i^S)^2]=\mu^2+\sigma^2$ which determine the values of $\lambda_{2,3}$.

%\newpage

%\bibliography{GLB}

\begin{thebibliography}{85}%
\makeatletter
\providecommand \@ifxundefined [1]{%
 \@ifx{#1\undefined}
}%
\providecommand \@ifnum [1]{%
 \ifnum #1\expandafter \@firstoftwo
 \else \expandafter \@secondoftwo
 \fi
}%
\providecommand \@ifx [1]{%
 \ifx #1\expandafter \@firstoftwo
 \else \expandafter \@secondoftwo
 \fi
}%
\providecommand \natexlab [1]{#1}%
\providecommand \enquote  [1]{``#1''}%
\providecommand \bibnamefont  [1]{#1}%
\providecommand \bibfnamefont [1]{#1}%
\providecommand \citenamefont [1]{#1}%
\providecommand \href@noop [0]{\@secondoftwo}%
\providecommand \href [0]{\begingroup \@sanitize@url \@href}%
\providecommand \@href[1]{\@@startlink{#1}\@@href}%
\providecommand \@@href[1]{\endgroup#1\@@endlink}%
\providecommand \@sanitize@url [0]{\catcode `\\12\catcode `\$12\catcode
  `\&12\catcode `\#12\catcode `\^12\catcode `\_12\catcode `\%12\relax}%
\providecommand \@@startlink[1]{}%
\providecommand \@@endlink[0]{}%
\providecommand \url  [0]{\begingroup\@sanitize@url \@url }%
\providecommand \@url [1]{\endgroup\@href {#1}{\urlprefix }}%
\providecommand \urlprefix  [0]{URL }%
\providecommand \Eprint [0]{\href }%
\providecommand \doibase [0]{https://doi.org/}%
\providecommand \selectlanguage [0]{\@gobble}%
\providecommand \bibinfo  [0]{\@secondoftwo}%
\providecommand \bibfield  [0]{\@secondoftwo}%
\providecommand \translation [1]{[#1]}%
\providecommand \BibitemOpen [0]{}%
\providecommand \bibitemStop [0]{}%
\providecommand \bibitemNoStop [0]{.\EOS\space}%
\providecommand \EOS [0]{\spacefactor3000\relax}%
\providecommand \BibitemShut  [1]{\csname bibitem#1\endcsname}%
\let\auto@bib@innerbib\@empty
%</preamble>
\bibitem [{\citenamefont {Parrondo}\ \emph {et~al.}(2015)\citenamefont
  {Parrondo}, \citenamefont {Horowitz},\ and\ \citenamefont
  {Sagawa}}]{Parrondo.15.NP}%
  \BibitemOpen
  \bibfield  {author} {\bibinfo {author} {\bibfnamefont {J.}~\bibnamefont
  {Parrondo}}, \bibinfo {author} {\bibfnamefont {J.}~\bibnamefont {Horowitz}},\
  and\ \bibinfo {author} {\bibfnamefont {T.}~\bibnamefont {Sagawa}},\
  }\bibfield  {title} {\bibinfo {title} {Thermodynamics of information},\
  }\href {http://dx.doi.org/10.1038/nphys3230} {\bibfield  {journal} {\bibinfo
  {journal} {Nat. Phys.}\ }\textbf {\bibinfo {volume} {11}},\ \bibinfo {pages}
  {131} (\bibinfo {year} {2015})}\BibitemShut {NoStop}%
\bibitem [{\citenamefont {Goold}\ \emph {et~al.}(2016)\citenamefont {Goold},
  \citenamefont {Huber}, \citenamefont {Riera}, \citenamefont {del Rio},\ and\
  \citenamefont {Skrzypczyk}}]{Goold.16.JPA}%
  \BibitemOpen
  \bibfield  {author} {\bibinfo {author} {\bibfnamefont {J.}~\bibnamefont
  {Goold}}, \bibinfo {author} {\bibfnamefont {M.}~\bibnamefont {Huber}},
  \bibinfo {author} {\bibfnamefont {A.}~\bibnamefont {Riera}}, \bibinfo
  {author} {\bibfnamefont {L.}~\bibnamefont {del Rio}},\ and\ \bibinfo {author}
  {\bibfnamefont {P.}~\bibnamefont {Skrzypczyk}},\ }\bibfield  {title}
  {\bibinfo {title} {The role of quantum information in thermodynamics—a
  topical review},\ }\href {http://stacks.iop.org/1751-8121/49/i=14/a=143001}
  {\bibfield  {journal} {\bibinfo  {journal} {J. Phys. A}\ }\textbf {\bibinfo
  {volume} {49}},\ \bibinfo {pages} {143001} (\bibinfo {year}
  {2016})}\BibitemShut {NoStop}%
\bibitem [{\citenamefont {Landauer}(1961)}]{Landauer.61.IBM}%
  \BibitemOpen
  \bibfield  {author} {\bibinfo {author} {\bibfnamefont {R.}~\bibnamefont
  {Landauer}},\ }\bibfield  {title} {\bibinfo {title} {Irreversibility and
  {H}eat {G}eneration in the {C}omputing {P}rocess},\ }\href
  {https://doi.org/10.1147/rd.53.0183} {\bibfield  {journal} {\bibinfo
  {journal} {IBM J. Res. Dev.}\ }\textbf {\bibinfo {volume} {5}},\ \bibinfo
  {pages} {183} (\bibinfo {year} {1961})}\BibitemShut {NoStop}%
\bibitem [{\citenamefont {Esposito}\ \emph {et~al.}(2010)\citenamefont
  {Esposito}, \citenamefont {Lindenberg},\ and\ \citenamefont {den
  Broeck}}]{Esposito.10.NJP}%
  \BibitemOpen
  \bibfield  {author} {\bibinfo {author} {\bibfnamefont {M.}~\bibnamefont
  {Esposito}}, \bibinfo {author} {\bibfnamefont {K.}~\bibnamefont
  {Lindenberg}},\ and\ \bibinfo {author} {\bibfnamefont {C.~V.}\ \bibnamefont
  {den Broeck}},\ }\bibfield  {title} {\bibinfo {title} {Entropy production as
  correlation between system and reservoir},\ }\href
  {https://doi.org/10.1088/1367-2630/12/1/013013} {\bibfield  {journal}
  {\bibinfo  {journal} {New J. Phys.}\ }\textbf {\bibinfo {volume} {12}},\
  \bibinfo {pages} {013013} (\bibinfo {year} {2010})}\BibitemShut {NoStop}%
\bibitem [{\citenamefont {Reeb}\ and\ \citenamefont
  {Wolf}(2014)}]{Reeb.14.NJP}%
  \BibitemOpen
  \bibfield  {author} {\bibinfo {author} {\bibfnamefont {D.}~\bibnamefont
  {Reeb}}\ and\ \bibinfo {author} {\bibfnamefont {M.}~\bibnamefont {Wolf}},\
  }\bibfield  {title} {\bibinfo {title} {An improved {L}andauer principle with
  finite-size corrections},\ }\href
  {https://doi.org/10.1088/1367-2630/16/10/103011} {\bibfield  {journal}
  {\bibinfo  {journal} {New J. Phys.}\ }\textbf {\bibinfo {volume} {16}},\
  \bibinfo {pages} {103011} (\bibinfo {year} {2014})}\BibitemShut {NoStop}%
\bibitem [{\citenamefont {Landi}\ and\ \citenamefont
  {Paternostro}(2021)}]{Landi.21.RMP}%
  \BibitemOpen
  \bibfield  {author} {\bibinfo {author} {\bibfnamefont {G.}~\bibnamefont
  {Landi}}\ and\ \bibinfo {author} {\bibfnamefont {M.}~\bibnamefont
  {Paternostro}},\ }\bibfield  {title} {\bibinfo {title} {Irreversible entropy
  production: {F}rom classical to quantum},\ }\href
  {https://doi.org/10.1103/RevModPhys.93.035008} {\bibfield  {journal}
  {\bibinfo  {journal} {Rev. Mod. Phys.}\ }\textbf {\bibinfo {volume} {93}},\
  \bibinfo {pages} {035008} (\bibinfo {year} {2021})}\BibitemShut {NoStop}%
\bibitem [{\citenamefont {Sagawa}\ and\ \citenamefont
  {Ueda}(2009)}]{Sagawa.09.PRL}%
  \BibitemOpen
  \bibfield  {author} {\bibinfo {author} {\bibfnamefont {T.}~\bibnamefont
  {Sagawa}}\ and\ \bibinfo {author} {\bibfnamefont {M.}~\bibnamefont {Ueda}},\
  }\bibfield  {title} {\bibinfo {title} {Minimal {E}nergy {C}ost for
  {T}hermodynamic {I}nformation {P}rocessing: {M}easurement and {I}nformation
  {E}rasure},\ }\href {https://doi.org/10.1103/PhysRevLett.102.250602}
  {\bibfield  {journal} {\bibinfo  {journal} {Phys. Rev. Lett.}\ }\textbf
  {\bibinfo {volume} {102}},\ \bibinfo {pages} {250602} (\bibinfo {year}
  {2009})}\BibitemShut {NoStop}%
\bibitem [{\citenamefont {Hilt}\ \emph {et~al.}(2011)\citenamefont {Hilt},
  \citenamefont {Shabbir}, \citenamefont {Anders},\ and\ \citenamefont
  {Lutz}}]{Hilt.11.PRE}%
  \BibitemOpen
  \bibfield  {author} {\bibinfo {author} {\bibfnamefont {S.}~\bibnamefont
  {Hilt}}, \bibinfo {author} {\bibfnamefont {S.}~\bibnamefont {Shabbir}},
  \bibinfo {author} {\bibfnamefont {J.}~\bibnamefont {Anders}},\ and\ \bibinfo
  {author} {\bibfnamefont {E.}~\bibnamefont {Lutz}},\ }\bibfield  {title}
  {\bibinfo {title} {Landauer's principle in the quantum regime},\ }\href
  {https://doi.org/10.1103/PhysRevE.83.030102} {\bibfield  {journal} {\bibinfo
  {journal} {Phys. Rev. E}\ }\textbf {\bibinfo {volume} {83}},\ \bibinfo
  {pages} {030102} (\bibinfo {year} {2011})}\BibitemShut {NoStop}%
\bibitem [{\citenamefont {Deffner}\ and\ \citenamefont
  {Jarzynski}(2013)}]{Deffner.13.PRX}%
  \BibitemOpen
  \bibfield  {author} {\bibinfo {author} {\bibfnamefont {S.}~\bibnamefont
  {Deffner}}\ and\ \bibinfo {author} {\bibfnamefont {C.}~\bibnamefont
  {Jarzynski}},\ }\bibfield  {title} {\bibinfo {title} {Information
  {P}rocessing and the {S}econd {L}aw of {T}hermodynamics: An {I}nclusive,
  {H}amiltonian {A}pproach},\ }\href
  {https://doi.org/10.1103/PhysRevX.3.041003} {\bibfield  {journal} {\bibinfo
  {journal} {Phys. Rev. X}\ }\textbf {\bibinfo {volume} {3}},\ \bibinfo {pages}
  {041003} (\bibinfo {year} {2013})}\BibitemShut {NoStop}%
\bibitem [{\citenamefont {Lorenzo}\ \emph {et~al.}(2015)\citenamefont
  {Lorenzo}, \citenamefont {McCloskey}, \citenamefont {Ciccarello},
  \citenamefont {Paternostro},\ and\ \citenamefont {Palma}}]{Lorenzo.15.PRL}%
  \BibitemOpen
  \bibfield  {author} {\bibinfo {author} {\bibfnamefont {S.}~\bibnamefont
  {Lorenzo}}, \bibinfo {author} {\bibfnamefont {R.}~\bibnamefont {McCloskey}},
  \bibinfo {author} {\bibfnamefont {F.}~\bibnamefont {Ciccarello}}, \bibinfo
  {author} {\bibfnamefont {M.}~\bibnamefont {Paternostro}},\ and\ \bibinfo
  {author} {\bibfnamefont {G.~M.}\ \bibnamefont {Palma}},\ }\bibfield  {title}
  {\bibinfo {title} {Landauer's {P}rinciple in {M}ultipartite {O}pen {Q}uantum
  {S}ystem {D}ynamics},\ }\href
  {https://doi.org/10.1103/PhysRevLett.115.120403} {\bibfield  {journal}
  {\bibinfo  {journal} {Phys. Rev. Lett.}\ }\textbf {\bibinfo {volume} {115}},\
  \bibinfo {pages} {120403} (\bibinfo {year} {2015})}\BibitemShut {NoStop}%
\bibitem [{\citenamefont {Dago}\ \emph {et~al.}(2021)\citenamefont {Dago},
  \citenamefont {Pereda}, \citenamefont {Barros}, \citenamefont {Ciliberto},\
  and\ \citenamefont {Bellon}}]{Dago.21.PRL}%
  \BibitemOpen
  \bibfield  {author} {\bibinfo {author} {\bibfnamefont {S.}~\bibnamefont
  {Dago}}, \bibinfo {author} {\bibfnamefont {J.}~\bibnamefont {Pereda}},
  \bibinfo {author} {\bibfnamefont {N.}~\bibnamefont {Barros}}, \bibinfo
  {author} {\bibfnamefont {S.}~\bibnamefont {Ciliberto}},\ and\ \bibinfo
  {author} {\bibfnamefont {L.}~\bibnamefont {Bellon}},\ }\bibfield  {title}
  {\bibinfo {title} {Information and {T}hermodynamics: {F}ast and {P}recise
  {A}pproach to {L}andauer's {B}ound in an {U}nderdamped {M}icromechanical
  {O}scillator},\ }\href {https://doi.org/10.1103/PhysRevLett.126.170601}
  {\bibfield  {journal} {\bibinfo  {journal} {Phys. Rev. Lett.}\ }\textbf
  {\bibinfo {volume} {126}},\ \bibinfo {pages} {170601} (\bibinfo {year}
  {2021})}\BibitemShut {NoStop}%
\bibitem [{\citenamefont {Riechers}\ and\ \citenamefont
  {Gu}(2021)}]{Riechers.21.PRA}%
  \BibitemOpen
  \bibfield  {author} {\bibinfo {author} {\bibfnamefont {P.}~\bibnamefont
  {Riechers}}\ and\ \bibinfo {author} {\bibfnamefont {M.}~\bibnamefont {Gu}},\
  }\bibfield  {title} {\bibinfo {title} {Impossibility of achieving
  {L}andauer's bound for almost every quantum state},\ }\href
  {https://doi.org/10.1103/PhysRevA.104.012214} {\bibfield  {journal} {\bibinfo
   {journal} {Phys. Rev. A}\ }\textbf {\bibinfo {volume} {104}},\ \bibinfo
  {pages} {012214} (\bibinfo {year} {2021})}\BibitemShut {NoStop}%
\bibitem [{\citenamefont {Goold}\ \emph {et~al.}(2015)\citenamefont {Goold},
  \citenamefont {Paternostro},\ and\ \citenamefont {Modi}}]{Goold.15.PRL}%
  \BibitemOpen
  \bibfield  {author} {\bibinfo {author} {\bibfnamefont {J.}~\bibnamefont
  {Goold}}, \bibinfo {author} {\bibfnamefont {M.}~\bibnamefont {Paternostro}},\
  and\ \bibinfo {author} {\bibfnamefont {K.}~\bibnamefont {Modi}},\ }\bibfield
  {title} {\bibinfo {title} {Nonequilibrium {Q}uantum {L}andauer {P}rinciple},\
  }\href {https://doi.org/10.1103/PhysRevLett.114.060602} {\bibfield  {journal}
  {\bibinfo  {journal} {Phys. Rev. Lett.}\ }\textbf {\bibinfo {volume} {114}},\
  \bibinfo {pages} {060602} (\bibinfo {year} {2015})}\BibitemShut {NoStop}%
\bibitem [{\citenamefont {Esposito}\ and\ \citenamefont {den
  Broeck}(2011)}]{Esposito.11.EPL}%
  \BibitemOpen
  \bibfield  {author} {\bibinfo {author} {\bibfnamefont {M.}~\bibnamefont
  {Esposito}}\ and\ \bibinfo {author} {\bibfnamefont {C.~V.}\ \bibnamefont {den
  Broeck}},\ }\bibfield  {title} {\bibinfo {title} {Second law and {L}andauer
  principle far from equilibrium},\ }\href
  {https://doi.org/10.1209/0295-5075/95/40004} {\bibfield  {journal} {\bibinfo
  {journal} {Europhys. Lett.}\ }\textbf {\bibinfo {volume} {95}},\ \bibinfo
  {pages} {40004} (\bibinfo {year} {2011})}\BibitemShut {NoStop}%
\bibitem [{\citenamefont {Campbell}\ \emph {et~al.}(2017)\citenamefont
  {Campbell}, \citenamefont {Guarnieri}, \citenamefont {Paternostro},\ and\
  \citenamefont {Vacchini}}]{Campbell.17.PRA}%
  \BibitemOpen
  \bibfield  {author} {\bibinfo {author} {\bibfnamefont {S.}~\bibnamefont
  {Campbell}}, \bibinfo {author} {\bibfnamefont {G.}~\bibnamefont {Guarnieri}},
  \bibinfo {author} {\bibfnamefont {M.}~\bibnamefont {Paternostro}},\ and\
  \bibinfo {author} {\bibfnamefont {B.}~\bibnamefont {Vacchini}},\ }\bibfield
  {title} {\bibinfo {title} {Nonequilibrium quantum bounds to {L}andauer's
  principle: Tightness and effectiveness},\ }\href
  {https://doi.org/10.1103/PhysRevA.96.042109} {\bibfield  {journal} {\bibinfo
  {journal} {Phys. Rev. A}\ }\textbf {\bibinfo {volume} {96}},\ \bibinfo
  {pages} {042109} (\bibinfo {year} {2017})}\BibitemShut {NoStop}%
\bibitem [{\citenamefont {Browne}\ \emph {et~al.}(2014)\citenamefont {Browne},
  \citenamefont {Garner}, \citenamefont {Dahlsten},\ and\ \citenamefont
  {Vedral}}]{Browne.14.PRL}%
  \BibitemOpen
  \bibfield  {author} {\bibinfo {author} {\bibfnamefont {C.}~\bibnamefont
  {Browne}}, \bibinfo {author} {\bibfnamefont {A.}~\bibnamefont {Garner}},
  \bibinfo {author} {\bibfnamefont {O.}~\bibnamefont {Dahlsten}},\ and\
  \bibinfo {author} {\bibfnamefont {V.}~\bibnamefont {Vedral}},\ }\bibfield
  {title} {\bibinfo {title} {Guaranteed energy-efficient bit reset in finite
  time},\ }\href {https://doi.org/10.1103/PhysRevLett.113.100603} {\bibfield
  {journal} {\bibinfo  {journal} {Phys. Rev. Lett.}\ }\textbf {\bibinfo
  {volume} {113}},\ \bibinfo {pages} {100603} (\bibinfo {year}
  {2014})}\BibitemShut {NoStop}%
\bibitem [{\citenamefont {Bera}\ \emph {et~al.}(2017)\citenamefont {Bera},
  \citenamefont {Riera}, \citenamefont {Lewenstein},\ and\ \citenamefont
  {Winter}}]{Bera.17.NC}%
  \BibitemOpen
  \bibfield  {author} {\bibinfo {author} {\bibfnamefont {M.}~\bibnamefont
  {Bera}}, \bibinfo {author} {\bibfnamefont {A.}~\bibnamefont {Riera}},
  \bibinfo {author} {\bibfnamefont {M.}~\bibnamefont {Lewenstein}},\ and\
  \bibinfo {author} {\bibfnamefont {A.}~\bibnamefont {Winter}},\ }\bibfield
  {title} {\bibinfo {title} {Generalized laws of thermodynamics in the presence
  of correlations},\ }\href {https://doi.org/10.1038/s41467-017-02370-x}
  {\bibfield  {journal} {\bibinfo  {journal} {Nat. Commun.}\ }\textbf {\bibinfo
  {volume} {8}},\ \bibinfo {pages} {2180} (\bibinfo {year} {2017})}\BibitemShut
  {NoStop}%
\bibitem [{\citenamefont {Miller}\ \emph {et~al.}(2020)\citenamefont {Miller},
  \citenamefont {Guarnieri}, \citenamefont {Mitchison},\ and\ \citenamefont
  {Goold}}]{Miller.20.PRL}%
  \BibitemOpen
  \bibfield  {author} {\bibinfo {author} {\bibfnamefont {H.}~\bibnamefont
  {Miller}}, \bibinfo {author} {\bibfnamefont {G.}~\bibnamefont {Guarnieri}},
  \bibinfo {author} {\bibfnamefont {M.}~\bibnamefont {Mitchison}},\ and\
  \bibinfo {author} {\bibfnamefont {J.}~\bibnamefont {Goold}},\ }\bibfield
  {title} {\bibinfo {title} {Quantum fluctuations hinder finite-time
  information erasure near the landauer limit},\ }\href
  {https://doi.org/10.1103/PhysRevLett.125.160602} {\bibfield  {journal}
  {\bibinfo  {journal} {Phys. Rev. Lett.}\ }\textbf {\bibinfo {volume} {125}},\
  \bibinfo {pages} {160602} (\bibinfo {year} {2020})}\BibitemShut {NoStop}%
\bibitem [{\citenamefont {Proesmans}\ \emph {et~al.}(2020)\citenamefont
  {Proesmans}, \citenamefont {Ehrich},\ and\ \citenamefont
  {Bechhoefer}}]{Proesmans.20.PRL}%
  \BibitemOpen
  \bibfield  {author} {\bibinfo {author} {\bibfnamefont {K.}~\bibnamefont
  {Proesmans}}, \bibinfo {author} {\bibfnamefont {J.}~\bibnamefont {Ehrich}},\
  and\ \bibinfo {author} {\bibfnamefont {J.}~\bibnamefont {Bechhoefer}},\
  }\bibfield  {title} {\bibinfo {title} {Finite-{T}ime {L}andauer
  {P}rinciple},\ }\href {https://doi.org/10.1103/PhysRevLett.125.100602}
  {\bibfield  {journal} {\bibinfo  {journal} {Phys. Rev. Lett.}\ }\textbf
  {\bibinfo {volume} {125}},\ \bibinfo {pages} {100602} (\bibinfo {year}
  {2020})}\BibitemShut {NoStop}%
\bibitem [{\citenamefont {Van~Vu}\ and\ \citenamefont
  {Saito}(2022)}]{Saito.22.PRL}%
  \BibitemOpen
  \bibfield  {author} {\bibinfo {author} {\bibfnamefont {T.}~\bibnamefont
  {Van~Vu}}\ and\ \bibinfo {author} {\bibfnamefont {K.}~\bibnamefont {Saito}},\
  }\bibfield  {title} {\bibinfo {title} {Finite-time quantum landauer principle
  and quantum coherence},\ }\href
  {https://doi.org/10.1103/PhysRevLett.128.010602} {\bibfield  {journal}
  {\bibinfo  {journal} {Phys. Rev. Lett.}\ }\textbf {\bibinfo {volume} {128}},\
  \bibinfo {pages} {010602} (\bibinfo {year} {2022})}\BibitemShut {NoStop}%
\bibitem [{\citenamefont {Lee}\ \emph {et~al.}(2022)\citenamefont {Lee},
  \citenamefont {Lee}, \citenamefont {Kwon},\ and\ \citenamefont
  {Park}}]{LeeJ.22.PRL}%
  \BibitemOpen
  \bibfield  {author} {\bibinfo {author} {\bibfnamefont {J.}~\bibnamefont
  {Lee}}, \bibinfo {author} {\bibfnamefont {S.}~\bibnamefont {Lee}}, \bibinfo
  {author} {\bibfnamefont {H.}~\bibnamefont {Kwon}},\ and\ \bibinfo {author}
  {\bibfnamefont {H.}~\bibnamefont {Park}},\ }\bibfield  {title} {\bibinfo
  {title} {Speed {L}imit for a {H}ighly {I}rreversible {P}rocess and {T}ight
  {F}inite-{T}ime {L}andauer's {B}ound},\ }\href
  {https://doi.org/10.1103/PhysRevLett.129.120603} {\bibfield  {journal}
  {\bibinfo  {journal} {Phys. Rev. Lett.}\ }\textbf {\bibinfo {volume} {129}},\
  \bibinfo {pages} {120603} (\bibinfo {year} {2022})}\BibitemShut {NoStop}%
\bibitem [{\citenamefont {Dago}\ and\ \citenamefont
  {Bellon}(2022)}]{Dago.22.PRL}%
  \BibitemOpen
  \bibfield  {author} {\bibinfo {author} {\bibfnamefont {S.}~\bibnamefont
  {Dago}}\ and\ \bibinfo {author} {\bibfnamefont {L.}~\bibnamefont {Bellon}},\
  }\bibfield  {title} {\bibinfo {title} {Dynamics of {I}nformation {E}rasure
  and {E}xtension of {L}andauer's {B}ound to {F}ast {P}rocesses},\ }\href
  {https://doi.org/10.1103/PhysRevLett.128.070604} {\bibfield  {journal}
  {\bibinfo  {journal} {Phys. Rev. Lett.}\ }\textbf {\bibinfo {volume} {128}},\
  \bibinfo {pages} {070604} (\bibinfo {year} {2022})}\BibitemShut {NoStop}%
\bibitem [{\citenamefont {Timpanaro}\ \emph {et~al.}(2020)\citenamefont
  {Timpanaro}, \citenamefont {Santos},\ and\ \citenamefont
  {Landi}}]{Timpanaro.20.PRL}%
  \BibitemOpen
  \bibfield  {author} {\bibinfo {author} {\bibfnamefont {A.}~\bibnamefont
  {Timpanaro}}, \bibinfo {author} {\bibfnamefont {J.}~\bibnamefont {Santos}},\
  and\ \bibinfo {author} {\bibfnamefont {G.}~\bibnamefont {Landi}},\ }\bibfield
   {title} {\bibinfo {title} {Landauer's principle at zero temperature},\
  }\href {https://doi.org/10.1103/PhysRevLett.124.240601} {\bibfield  {journal}
  {\bibinfo  {journal} {Phys. Rev. Lett.}\ }\textbf {\bibinfo {volume} {124}},\
  \bibinfo {pages} {240601} (\bibinfo {year} {2020})}\BibitemShut {NoStop}%
\bibitem [{\citenamefont {Hsieh}(2025)}]{Hsieh.25.PRL}%
  \BibitemOpen
  \bibfield  {author} {\bibinfo {author} {\bibfnamefont {C.-Y.}\ \bibnamefont
  {Hsieh}},\ }\bibfield  {title} {\bibinfo {title} {Dynamical {L}andauer
  principle: Quantifying information transmission by thermodynamics},\ }\href
  {https://doi.org/10.1103/PhysRevLett.134.050404} {\bibfield  {journal}
  {\bibinfo  {journal} {Phys. Rev. Lett.}\ }\textbf {\bibinfo {volume} {134}},\
  \bibinfo {pages} {050404} (\bibinfo {year} {2025})}\BibitemShut {NoStop}%
\bibitem [{\citenamefont {Liu}\ and\ \citenamefont {Nie}(2023)}]{Liu.23.PRAa}%
  \BibitemOpen
  \bibfield  {author} {\bibinfo {author} {\bibfnamefont {J.}~\bibnamefont
  {Liu}}\ and\ \bibinfo {author} {\bibfnamefont {H.}~\bibnamefont {Nie}},\
  }\bibfield  {title} {\bibinfo {title} {Universal landauer-like inequality
  from the first law of thermodynamics},\ }\href
  {https://doi.org/10.1103/PhysRevA.108.L040203} {\bibfield  {journal}
  {\bibinfo  {journal} {Phys. Rev. A}\ }\textbf {\bibinfo {volume} {108}},\
  \bibinfo {pages} {L040203} (\bibinfo {year} {2023})}\BibitemShut {NoStop}%
\bibitem [{\citenamefont {Liu}\ \emph {et~al.}(2024)\citenamefont {Liu},
  \citenamefont {Lu}, \citenamefont {Wang},\ and\ \citenamefont
  {Jiang}}]{LiuJ.24.PRR}%
  \BibitemOpen
  \bibfield  {author} {\bibinfo {author} {\bibfnamefont {J.}~\bibnamefont
  {Liu}}, \bibinfo {author} {\bibfnamefont {J.}~\bibnamefont {Lu}}, \bibinfo
  {author} {\bibfnamefont {C.}~\bibnamefont {Wang}},\ and\ \bibinfo {author}
  {\bibfnamefont {J.-H.}\ \bibnamefont {Jiang}},\ }\bibfield  {title} {\bibinfo
  {title} {Inferring general links between energetics and information with
  unknown environment},\ }\href
  {https://doi.org/10.1103/PhysRevResearch.6.033202} {\bibfield  {journal}
  {\bibinfo  {journal} {Phys. Rev. Res.}\ }\textbf {\bibinfo {volume} {6}},\
  \bibinfo {pages} {033202} (\bibinfo {year} {2024})}\BibitemShut {NoStop}%
\bibitem [{\citenamefont {Peterson}\ \emph {et~al.}(2016)\citenamefont
  {Peterson}, \citenamefont {Sarthour}, \citenamefont {Souza}, \citenamefont
  {Oliveira}, \citenamefont {Goold}, \citenamefont {Modi}, \citenamefont
  {Soares-Pinto},\ and\ \citenamefont {Céleri}}]{Peterson.16.PRSA}%
  \BibitemOpen
  \bibfield  {author} {\bibinfo {author} {\bibfnamefont {J.}~\bibnamefont
  {Peterson}}, \bibinfo {author} {\bibfnamefont {R.}~\bibnamefont {Sarthour}},
  \bibinfo {author} {\bibfnamefont {A.}~\bibnamefont {Souza}}, \bibinfo
  {author} {\bibfnamefont {I.}~\bibnamefont {Oliveira}}, \bibinfo {author}
  {\bibfnamefont {J.}~\bibnamefont {Goold}}, \bibinfo {author} {\bibfnamefont
  {K.}~\bibnamefont {Modi}}, \bibinfo {author} {\bibfnamefont {D.}~\bibnamefont
  {Soares-Pinto}},\ and\ \bibinfo {author} {\bibfnamefont {L.}~\bibnamefont
  {Céleri}},\ }\bibfield  {title} {\bibinfo {title} {Experimental
  demonstration of information to energy conversion in a quantum system at the
  {L}andauer limit},\ }\href {https://doi.org/10.1098/rspa.2015.0813}
  {\bibfield  {journal} {\bibinfo  {journal} {Proc. R. Soc. A}\ }\textbf
  {\bibinfo {volume} {472}},\ \bibinfo {pages} {20150813} (\bibinfo {year}
  {2016})}\BibitemShut {NoStop}%
\bibitem [{\citenamefont {Yan}\ \emph {et~al.}(2018)\citenamefont {Yan},
  \citenamefont {Xiong}, \citenamefont {Rehan}, \citenamefont {Zhou},
  \citenamefont {Liang}, \citenamefont {Chen}, \citenamefont {Zhang},
  \citenamefont {Yang}, \citenamefont {Ma},\ and\ \citenamefont
  {Feng}}]{Yan.18.PRL}%
  \BibitemOpen
  \bibfield  {author} {\bibinfo {author} {\bibfnamefont {L.}~\bibnamefont
  {Yan}}, \bibinfo {author} {\bibfnamefont {T.}~\bibnamefont {Xiong}}, \bibinfo
  {author} {\bibfnamefont {K.}~\bibnamefont {Rehan}}, \bibinfo {author}
  {\bibfnamefont {F.}~\bibnamefont {Zhou}}, \bibinfo {author} {\bibfnamefont
  {D.}~\bibnamefont {Liang}}, \bibinfo {author} {\bibfnamefont
  {L.}~\bibnamefont {Chen}}, \bibinfo {author} {\bibfnamefont {J.}~\bibnamefont
  {Zhang}}, \bibinfo {author} {\bibfnamefont {W.}~\bibnamefont {Yang}},
  \bibinfo {author} {\bibfnamefont {Z.}~\bibnamefont {Ma}},\ and\ \bibinfo
  {author} {\bibfnamefont {M.}~\bibnamefont {Feng}},\ }\bibfield  {title}
  {\bibinfo {title} {Single-{A}tom {D}emonstration of the {Q}uantum {L}andauer
  {P}rinciple},\ }\href {https://doi.org/10.1103/PhysRevLett.120.210601}
  {\bibfield  {journal} {\bibinfo  {journal} {Phys. Rev. Lett.}\ }\textbf
  {\bibinfo {volume} {120}},\ \bibinfo {pages} {210601} (\bibinfo {year}
  {2018})}\BibitemShut {NoStop}%
\bibitem [{\citenamefont {Gaudenzi}\ \emph {et~al.}(2018)\citenamefont
  {Gaudenzi}, \citenamefont {Burzurí}, \citenamefont {Maegawa}, \citenamefont
  {van~der Zant},\ and\ \citenamefont {Luis}}]{Gaudenzi.18.NP}%
  \BibitemOpen
  \bibfield  {author} {\bibinfo {author} {\bibfnamefont {R.}~\bibnamefont
  {Gaudenzi}}, \bibinfo {author} {\bibfnamefont {E.}~\bibnamefont {Burzurí}},
  \bibinfo {author} {\bibfnamefont {S.}~\bibnamefont {Maegawa}}, \bibinfo
  {author} {\bibfnamefont {H.}~\bibnamefont {van~der Zant}},\ and\ \bibinfo
  {author} {\bibfnamefont {F.}~\bibnamefont {Luis}},\ }\bibfield  {title}
  {\bibinfo {title} {Quantum {L}andauer erasure with a molecular nanomagnet},\
  }\href {https://doi.org/10.1038/s41567-018-0070-7} {\bibfield  {journal}
  {\bibinfo  {journal} {Nat. Phys.}\ }\textbf {\bibinfo {volume} {14}},\
  \bibinfo {pages} {565} (\bibinfo {year} {2018})}\BibitemShut {NoStop}%
\bibitem [{\citenamefont {Aimet}\ \emph {et~al.}(2025)\citenamefont {Aimet},
  \citenamefont {Tajik}, \citenamefont {Tournaire}, \citenamefont
  {Sch{\"u}ttelkopf}, \citenamefont {Sabino}, \citenamefont {Sotiriadis},
  \citenamefont {Guarnieri}, \citenamefont {Schmiedmayer},\ and\ \citenamefont
  {Eisert}}]{Aimet.25.NP}%
  \BibitemOpen
  \bibfield  {author} {\bibinfo {author} {\bibfnamefont {S.}~\bibnamefont
  {Aimet}}, \bibinfo {author} {\bibfnamefont {M.}~\bibnamefont {Tajik}},
  \bibinfo {author} {\bibfnamefont {G.}~\bibnamefont {Tournaire}}, \bibinfo
  {author} {\bibfnamefont {P.}~\bibnamefont {Sch{\"u}ttelkopf}}, \bibinfo
  {author} {\bibfnamefont {J.}~\bibnamefont {Sabino}}, \bibinfo {author}
  {\bibfnamefont {S.}~\bibnamefont {Sotiriadis}}, \bibinfo {author}
  {\bibfnamefont {G.}~\bibnamefont {Guarnieri}}, \bibinfo {author}
  {\bibfnamefont {J.}~\bibnamefont {Schmiedmayer}},\ and\ \bibinfo {author}
  {\bibfnamefont {J.}~\bibnamefont {Eisert}},\ }\bibfield  {title} {\bibinfo
  {title} {Experimentally probing {L}andauer's principle in the quantum
  many-body regime},\ }\href {https://doi.org/10.1038/s41567-025-02930-9}
  {\bibfield  {journal} {\bibinfo  {journal} {Nat. Phys.}\ }\textbf {\bibinfo
  {volume} {21}},\ \bibinfo {pages} {1326} (\bibinfo {year}
  {2025})}\BibitemShut {NoStop}%
\bibitem [{\citenamefont {Chattopadhyay}\ \emph {et~al.}(2025)\citenamefont
  {Chattopadhyay}, \citenamefont {Misra}, \citenamefont {Pandit},\ and\
  \citenamefont {Paul}}]{Chattopadhyay.25.RPP}%
  \BibitemOpen
  \bibfield  {author} {\bibinfo {author} {\bibfnamefont {P.}~\bibnamefont
  {Chattopadhyay}}, \bibinfo {author} {\bibfnamefont {A.}~\bibnamefont
  {Misra}}, \bibinfo {author} {\bibfnamefont {T.}~\bibnamefont {Pandit}},\ and\
  \bibinfo {author} {\bibfnamefont {G.}~\bibnamefont {Paul}},\ }\bibfield
  {title} {\bibinfo {title} {Landauer principle and thermodynamics of
  computation},\ }\href {https://doi.org/10.1088/1361-6633/add6b3} {\bibfield
  {journal} {\bibinfo  {journal} {Rep. Prog. Phys.}\ }\textbf {\bibinfo
  {volume} {88}},\ \bibinfo {pages} {086001} (\bibinfo {year}
  {2025})}\BibitemShut {NoStop}%
\bibitem [{\citenamefont {Langen}\ \emph {et~al.}(2015)\citenamefont {Langen},
  \citenamefont {Erne}, \citenamefont {Geiger}, \citenamefont {Rauer},
  \citenamefont {Schweigler}, \citenamefont {Kuhnert}, \citenamefont
  {Rohringer}, \citenamefont {Mazets}, \citenamefont {Gasenzer},\ and\
  \citenamefont {Schmiedmayer}}]{Langen.S.15}%
  \BibitemOpen
  \bibfield  {author} {\bibinfo {author} {\bibfnamefont {T.}~\bibnamefont
  {Langen}}, \bibinfo {author} {\bibfnamefont {S.}~\bibnamefont {Erne}},
  \bibinfo {author} {\bibfnamefont {R.}~\bibnamefont {Geiger}}, \bibinfo
  {author} {\bibfnamefont {B.}~\bibnamefont {Rauer}}, \bibinfo {author}
  {\bibfnamefont {T.}~\bibnamefont {Schweigler}}, \bibinfo {author}
  {\bibfnamefont {M.}~\bibnamefont {Kuhnert}}, \bibinfo {author} {\bibfnamefont
  {W.}~\bibnamefont {Rohringer}}, \bibinfo {author} {\bibfnamefont
  {I.}~\bibnamefont {Mazets}}, \bibinfo {author} {\bibfnamefont
  {T.}~\bibnamefont {Gasenzer}},\ and\ \bibinfo {author} {\bibfnamefont
  {J.}~\bibnamefont {Schmiedmayer}},\ }\bibfield  {title} {\bibinfo {title}
  {Experimental observation of a generalized gibbs ensemble},\ }\href
  {https://doi.org/10.1126/science.1257026} {\bibfield  {journal} {\bibinfo
  {journal} {Science}\ }\textbf {\bibinfo {volume} {348}},\ \bibinfo {pages}
  {207} (\bibinfo {year} {2015})}\BibitemShut {NoStop}%
\bibitem [{\citenamefont {\'Alvarez}\ \emph {et~al.}(2015)\citenamefont
  {\'Alvarez}, \citenamefont {Suter},\ and\ \citenamefont
  {Kaiser}}]{Alvarez.15.S}%
  \BibitemOpen
  \bibfield  {author} {\bibinfo {author} {\bibfnamefont {G.}~\bibnamefont
  {\'Alvarez}}, \bibinfo {author} {\bibfnamefont {D.}~\bibnamefont {Suter}},\
  and\ \bibinfo {author} {\bibfnamefont {R.}~\bibnamefont {Kaiser}},\
  }\bibfield  {title} {\bibinfo {title} {Localization-delocalization transition
  in the dynamics of dipolar-coupled nuclear spins},\ }\href
  {https://doi.org/10.1126/science.1261160} {\bibfield  {journal} {\bibinfo
  {journal} {Science}\ }\textbf {\bibinfo {volume} {349}},\ \bibinfo {pages}
  {846} (\bibinfo {year} {2015})}\BibitemShut {NoStop}%
\bibitem [{\citenamefont {Degen}\ \emph {et~al.}(2017)\citenamefont {Degen},
  \citenamefont {Reinhard},\ and\ \citenamefont {Cappellaro}}]{Degen.17.RMP}%
  \BibitemOpen
  \bibfield  {author} {\bibinfo {author} {\bibfnamefont {C.}~\bibnamefont
  {Degen}}, \bibinfo {author} {\bibfnamefont {F.}~\bibnamefont {Reinhard}},\
  and\ \bibinfo {author} {\bibfnamefont {P.}~\bibnamefont {Cappellaro}},\
  }\bibfield  {title} {\bibinfo {title} {Quantum sensing},\ }\href
  {https://doi.org/10.1103/RevModPhys.89.035002} {\bibfield  {journal}
  {\bibinfo  {journal} {Rev. Mod. Phys.}\ }\textbf {\bibinfo {volume} {89}},\
  \bibinfo {pages} {035002} (\bibinfo {year} {2017})}\BibitemShut {NoStop}%
\bibitem [{\citenamefont {Yang}\ \emph {et~al.}(2016)\citenamefont {Yang},
  \citenamefont {Ma},\ and\ \citenamefont {Liu}}]{Yang.17.RPP}%
  \BibitemOpen
  \bibfield  {author} {\bibinfo {author} {\bibfnamefont {W.}~\bibnamefont
  {Yang}}, \bibinfo {author} {\bibfnamefont {W.-L.}\ \bibnamefont {Ma}},\ and\
  \bibinfo {author} {\bibfnamefont {R.-B.}\ \bibnamefont {Liu}},\ }\bibfield
  {title} {\bibinfo {title} {Quantum many-body theory for electron spin
  decoherence in nanoscale nuclear spin baths},\ }\href
  {https://doi.org/10.1088/0034-4885/80/1/016001} {\bibfield  {journal}
  {\bibinfo  {journal} {Rep. Prog. Phys.}\ }\textbf {\bibinfo {volume} {80}},\
  \bibinfo {pages} {016001} (\bibinfo {year} {2016})}\BibitemShut {NoStop}%
\bibitem [{\citenamefont {Wang}\ \emph {et~al.}(2019)\citenamefont {Wang},
  \citenamefont {Chen}, \citenamefont {Peng}, \citenamefont {Wrachtrup},\ and\
  \citenamefont {Liu}}]{WangP.19.PRL}%
  \BibitemOpen
  \bibfield  {author} {\bibinfo {author} {\bibfnamefont {P.}~\bibnamefont
  {Wang}}, \bibinfo {author} {\bibfnamefont {C.}~\bibnamefont {Chen}}, \bibinfo
  {author} {\bibfnamefont {X.}~\bibnamefont {Peng}}, \bibinfo {author}
  {\bibfnamefont {J.}~\bibnamefont {Wrachtrup}},\ and\ \bibinfo {author}
  {\bibfnamefont {R.-B.}\ \bibnamefont {Liu}},\ }\bibfield  {title} {\bibinfo
  {title} {Characterization of arbitrary-order correlations in quantum baths by
  weak measurement},\ }\href {https://doi.org/10.1103/PhysRevLett.123.050603}
  {\bibfield  {journal} {\bibinfo  {journal} {Phys. Rev. Lett.}\ }\textbf
  {\bibinfo {volume} {123}},\ \bibinfo {pages} {050603} (\bibinfo {year}
  {2019})}\BibitemShut {NoStop}%
\bibitem [{\citenamefont {Abanin}\ \emph {et~al.}(2019)\citenamefont {Abanin},
  \citenamefont {Altman}, \citenamefont {Bloch},\ and\ \citenamefont
  {Serbyn}}]{Abanin.19.RMP}%
  \BibitemOpen
  \bibfield  {author} {\bibinfo {author} {\bibfnamefont {D.}~\bibnamefont
  {Abanin}}, \bibinfo {author} {\bibfnamefont {E.}~\bibnamefont {Altman}},
  \bibinfo {author} {\bibfnamefont {I.}~\bibnamefont {Bloch}},\ and\ \bibinfo
  {author} {\bibfnamefont {M.}~\bibnamefont {Serbyn}},\ }\bibfield  {title}
  {\bibinfo {title} {Colloquium: Many-body localization, thermalization, and
  entanglement},\ }\href {https://doi.org/10.1103/RevModPhys.91.021001}
  {\bibfield  {journal} {\bibinfo  {journal} {Rev. Mod. Phys.}\ }\textbf
  {\bibinfo {volume} {91}},\ \bibinfo {pages} {021001} (\bibinfo {year}
  {2019})}\BibitemShut {NoStop}%
\bibitem [{\citenamefont {von L\"upke}\ \emph {et~al.}(2020)\citenamefont {von
  L\"upke}, \citenamefont {Beaudoin}, \citenamefont {Norris}, \citenamefont
  {Sung}, \citenamefont {Winik}, \citenamefont {Qiu}, \citenamefont
  {Kjaergaard}, \citenamefont {Kim}, \citenamefont {Yoder}, \citenamefont
  {Gustavsson}, \citenamefont {Viola},\ and\ \citenamefont
  {Oliver}}]{Lupke.20.PRXQ}%
  \BibitemOpen
  \bibfield  {author} {\bibinfo {author} {\bibfnamefont {U.}~\bibnamefont {von
  L\"upke}}, \bibinfo {author} {\bibfnamefont {F.}~\bibnamefont {Beaudoin}},
  \bibinfo {author} {\bibfnamefont {L.}~\bibnamefont {Norris}}, \bibinfo
  {author} {\bibfnamefont {Y.}~\bibnamefont {Sung}}, \bibinfo {author}
  {\bibfnamefont {R.}~\bibnamefont {Winik}}, \bibinfo {author} {\bibfnamefont
  {J.}~\bibnamefont {Qiu}}, \bibinfo {author} {\bibfnamefont {M.}~\bibnamefont
  {Kjaergaard}}, \bibinfo {author} {\bibfnamefont {D.}~\bibnamefont {Kim}},
  \bibinfo {author} {\bibfnamefont {J.}~\bibnamefont {Yoder}}, \bibinfo
  {author} {\bibfnamefont {S.}~\bibnamefont {Gustavsson}}, \bibinfo {author}
  {\bibfnamefont {L.}~\bibnamefont {Viola}},\ and\ \bibinfo {author}
  {\bibfnamefont {W.}~\bibnamefont {Oliver}},\ }\bibfield  {title} {\bibinfo
  {title} {Two-qubit spectroscopy of spatiotemporally correlated quantum noise
  in superconducting qubits},\ }\href
  {https://doi.org/10.1103/PRXQuantum.1.010305} {\bibfield  {journal} {\bibinfo
   {journal} {PRX Quantum}\ }\textbf {\bibinfo {volume} {1}},\ \bibinfo {pages}
  {010305} (\bibinfo {year} {2020})}\BibitemShut {NoStop}%
\bibitem [{\citenamefont {Jackson}\ \emph {et~al.}(2021)\citenamefont
  {Jackson}, \citenamefont {Gangloff}, \citenamefont {Bodey}, \citenamefont
  {Zaporski}, \citenamefont {Bachorz}, \citenamefont {Clarke}, \citenamefont
  {Hugues}, \citenamefont {Le~Gall},\ and\ \citenamefont
  {Atat\"ure}}]{Jackson.21.NP}%
  \BibitemOpen
  \bibfield  {author} {\bibinfo {author} {\bibfnamefont {D.}~\bibnamefont
  {Jackson}}, \bibinfo {author} {\bibfnamefont {D.}~\bibnamefont {Gangloff}},
  \bibinfo {author} {\bibfnamefont {J.}~\bibnamefont {Bodey}}, \bibinfo
  {author} {\bibfnamefont {L.}~\bibnamefont {Zaporski}}, \bibinfo {author}
  {\bibfnamefont {C.}~\bibnamefont {Bachorz}}, \bibinfo {author} {\bibfnamefont
  {E.}~\bibnamefont {Clarke}}, \bibinfo {author} {\bibfnamefont
  {M.}~\bibnamefont {Hugues}}, \bibinfo {author} {\bibfnamefont
  {C.}~\bibnamefont {Le~Gall}},\ and\ \bibinfo {author} {\bibfnamefont
  {M.}~\bibnamefont {Atat\"ure}},\ }\bibfield  {title} {\bibinfo {title}
  {Quantum sensing of a coherent single spin excitation in a nuclear
  ensemble},\ }\href {https://doi.org/10.1038/s41567-020-01161-4} {\bibfield
  {journal} {\bibinfo  {journal} {Nat. Phys.}\ }\textbf {\bibinfo {volume}
  {17}},\ \bibinfo {pages} {585} (\bibinfo {year} {2021})}\BibitemShut
  {NoStop}%
\bibitem [{\citenamefont {Kuffer}\ \emph {et~al.}(2022)\citenamefont {Kuffer},
  \citenamefont {Zwick},\ and\ \citenamefont {\'Alvarez}}]{Kuffer.22.PRXQ}%
  \BibitemOpen
  \bibfield  {author} {\bibinfo {author} {\bibfnamefont {M.}~\bibnamefont
  {Kuffer}}, \bibinfo {author} {\bibfnamefont {A.}~\bibnamefont {Zwick}},\ and\
  \bibinfo {author} {\bibfnamefont {G.}~\bibnamefont {\'Alvarez}},\ }\bibfield
  {title} {\bibinfo {title} {Path integral framework for characterizing and
  controlling decoherence induced by nonstationary environments on a quantum
  probe},\ }\href {https://doi.org/10.1103/PRXQuantum.3.020321} {\bibfield
  {journal} {\bibinfo  {journal} {PRX Quantum}\ }\textbf {\bibinfo {volume}
  {3}},\ \bibinfo {pages} {020321} (\bibinfo {year} {2022})}\BibitemShut
  {NoStop}%
\bibitem [{\citenamefont {Kuffer}\ \emph {et~al.}(2025)\citenamefont {Kuffer},
  \citenamefont {Zwick},\ and\ \citenamefont {\'Alvarez}}]{Kuffer.25.PRXQ}%
  \BibitemOpen
  \bibfield  {author} {\bibinfo {author} {\bibfnamefont {M.}~\bibnamefont
  {Kuffer}}, \bibinfo {author} {\bibfnamefont {A.}~\bibnamefont {Zwick}},\ and\
  \bibinfo {author} {\bibfnamefont {G.}~\bibnamefont {\'Alvarez}},\ }\bibfield
  {title} {\bibinfo {title} {Sensing out-of-equilibrium and quantum
  non-gaussian environments via induced time-reversal symmetry breaking on the
  quantum-probe dynamics},\ }\href
  {https://doi.org/10.1103/PRXQuantum.6.020320} {\bibfield  {journal} {\bibinfo
   {journal} {PRX Quantum}\ }\textbf {\bibinfo {volume} {6}},\ \bibinfo {pages}
  {020320} (\bibinfo {year} {2025})}\BibitemShut {NoStop}%
\bibitem [{\citenamefont {Wise}\ \emph {et~al.}(2021)\citenamefont {Wise},
  \citenamefont {Morton},\ and\ \citenamefont {Dhomkar}}]{Wise.21.PRXQ}%
  \BibitemOpen
  \bibfield  {author} {\bibinfo {author} {\bibfnamefont {D.}~\bibnamefont
  {Wise}}, \bibinfo {author} {\bibfnamefont {J.}~\bibnamefont {Morton}},\ and\
  \bibinfo {author} {\bibfnamefont {S.}~\bibnamefont {Dhomkar}},\ }\bibfield
  {title} {\bibinfo {title} {Using deep learning to understand and mitigate the
  qubit noise environment},\ }\href
  {https://doi.org/10.1103/PRXQuantum.2.010316} {\bibfield  {journal} {\bibinfo
   {journal} {PRX Quantum}\ }\textbf {\bibinfo {volume} {2}},\ \bibinfo {pages}
  {010316} (\bibinfo {year} {2021})}\BibitemShut {NoStop}%
\bibitem [{\citenamefont {Martina}\ \emph {et~al.}(2023)\citenamefont
  {Martina}, \citenamefont {Gherardini},\ and\ \citenamefont
  {Caruso}}]{Martina.23.PS}%
  \BibitemOpen
  \bibfield  {author} {\bibinfo {author} {\bibfnamefont {S.}~\bibnamefont
  {Martina}}, \bibinfo {author} {\bibfnamefont {S.}~\bibnamefont
  {Gherardini}},\ and\ \bibinfo {author} {\bibfnamefont {F.}~\bibnamefont
  {Caruso}},\ }\bibfield  {title} {\bibinfo {title} {Machine learning
  classification of non-markovian noise disturbing quantum dynamics},\ }\href
  {https://doi.org/10.1088/1402-4896/acb39b} {\bibfield  {journal} {\bibinfo
  {journal} {Phys. Scr.}\ }\textbf {\bibinfo {volume} {98}},\ \bibinfo {pages}
  {035104} (\bibinfo {year} {2023})}\BibitemShut {NoStop}%
\bibitem [{\citenamefont {Chen}\ \emph {et~al.}(2019)\citenamefont {Chen},
  \citenamefont {Lo}, \citenamefont {Gneiting}, \citenamefont {Bae},
  \citenamefont {Chen},\ and\ \citenamefont {Nori}}]{Chen.19.NC}%
  \BibitemOpen
  \bibfield  {author} {\bibinfo {author} {\bibfnamefont {H.}~\bibnamefont
  {Chen}}, \bibinfo {author} {\bibfnamefont {P.}~\bibnamefont {Lo}}, \bibinfo
  {author} {\bibfnamefont {C.}~\bibnamefont {Gneiting}}, \bibinfo {author}
  {\bibfnamefont {J.}~\bibnamefont {Bae}}, \bibinfo {author} {\bibfnamefont
  {Y.}~\bibnamefont {Chen}},\ and\ \bibinfo {author} {\bibfnamefont
  {F.}~\bibnamefont {Nori}},\ }\bibfield  {title} {\bibinfo {title}
  {Quantifying the nonclassicality of pure dephasing},\ }\href
  {https://doi.org/10.1038/s41467-019-11502-4} {\bibfield  {journal} {\bibinfo
  {journal} {Nat. Commun.}\ }\textbf {\bibinfo {volume} {10}},\ \bibinfo
  {pages} {3794} (\bibinfo {year} {2019})}\BibitemShut {NoStop}%
\bibitem [{\citenamefont {Spiecker}\ \emph {et~al.}(2023)\citenamefont
  {Spiecker}, \citenamefont {Paluch}, \citenamefont {Gosling}, \citenamefont
  {Drucker}, \citenamefont {Matityahu}, \citenamefont {Gusenkova},
  \citenamefont {G{\"u}nzler}, \citenamefont {Rieger}, \citenamefont
  {Takmakov}, \citenamefont {Valenti}, \citenamefont {Winkel}, \citenamefont
  {Gebauer}, \citenamefont {Sander}, \citenamefont {Catelani}, \citenamefont
  {Shnirman}, \citenamefont {Ustinov}, \citenamefont {Wernsdorfer},
  \citenamefont {Cohen},\ and\ \citenamefont {Pop}}]{Spiecker.22.NP}%
  \BibitemOpen
  \bibfield  {author} {\bibinfo {author} {\bibfnamefont {M.}~\bibnamefont
  {Spiecker}}, \bibinfo {author} {\bibfnamefont {P.}~\bibnamefont {Paluch}},
  \bibinfo {author} {\bibfnamefont {N.}~\bibnamefont {Gosling}}, \bibinfo
  {author} {\bibfnamefont {N.}~\bibnamefont {Drucker}}, \bibinfo {author}
  {\bibfnamefont {S.}~\bibnamefont {Matityahu}}, \bibinfo {author}
  {\bibfnamefont {D.}~\bibnamefont {Gusenkova}}, \bibinfo {author}
  {\bibfnamefont {S.}~\bibnamefont {G{\"u}nzler}}, \bibinfo {author}
  {\bibfnamefont {D.}~\bibnamefont {Rieger}}, \bibinfo {author} {\bibfnamefont
  {I.}~\bibnamefont {Takmakov}}, \bibinfo {author} {\bibfnamefont
  {F.}~\bibnamefont {Valenti}}, \bibinfo {author} {\bibfnamefont
  {P.}~\bibnamefont {Winkel}}, \bibinfo {author} {\bibfnamefont
  {R.}~\bibnamefont {Gebauer}}, \bibinfo {author} {\bibfnamefont
  {O.}~\bibnamefont {Sander}}, \bibinfo {author} {\bibfnamefont
  {G.}~\bibnamefont {Catelani}}, \bibinfo {author} {\bibfnamefont
  {A.}~\bibnamefont {Shnirman}}, \bibinfo {author} {\bibfnamefont
  {A.}~\bibnamefont {Ustinov}}, \bibinfo {author} {\bibfnamefont
  {W.}~\bibnamefont {Wernsdorfer}}, \bibinfo {author} {\bibfnamefont
  {Y.}~\bibnamefont {Cohen}},\ and\ \bibinfo {author} {\bibfnamefont
  {I.}~\bibnamefont {Pop}},\ }\bibfield  {title} {\bibinfo {title} {Two-level
  system hyperpolarization using a quantum szilard engine},\ }\href
  {https://doi.org/10.1038/s41567-023-02082-8} {\bibfield  {journal} {\bibinfo
  {journal} {Nat. Phys.}\ }\textbf {\bibinfo {volume} {19}},\ \bibinfo {pages}
  {1320} (\bibinfo {year} {2023})}\BibitemShut {NoStop}%
\bibitem [{\citenamefont {Vaccaro}\ and\ \citenamefont
  {Barnett}(2011)}]{Vaccaro.11.PRSA}%
  \BibitemOpen
  \bibfield  {author} {\bibinfo {author} {\bibfnamefont {J.}~\bibnamefont
  {Vaccaro}}\ and\ \bibinfo {author} {\bibfnamefont {S.}~\bibnamefont
  {Barnett}},\ }\bibfield  {title} {\bibinfo {title} {Information erasure
  without an energy cost},\ }\href {https://doi.org/10.1098/rspa.2010.0577}
  {\bibfield  {journal} {\bibinfo  {journal} {Proc. R. Soc. A}\ }\textbf
  {\bibinfo {volume} {467}},\ \bibinfo {pages} {1770} (\bibinfo {year}
  {2011})}\BibitemShut {NoStop}%
\bibitem [{\citenamefont {Barnett}\ and\ \citenamefont
  {Vaccaro}(2013)}]{Barnett.13.E}%
  \BibitemOpen
  \bibfield  {author} {\bibinfo {author} {\bibfnamefont {S.}~\bibnamefont
  {Barnett}}\ and\ \bibinfo {author} {\bibfnamefont {J.}~\bibnamefont
  {Vaccaro}},\ }\bibfield  {title} {\bibinfo {title} {Beyond landauer
  erasure},\ }\href {https://doi.org/10.3390/e15114956} {\bibfield  {journal}
  {\bibinfo  {journal} {Entropy}\ }\textbf {\bibinfo {volume} {15}},\ \bibinfo
  {pages} {4956} (\bibinfo {year} {2013})}\BibitemShut {NoStop}%
\bibitem [{\citenamefont {Lostaglio}\ \emph {et~al.}(2017)\citenamefont
  {Lostaglio}, \citenamefont {Jennings},\ and\ \citenamefont
  {Rudolph}}]{Lostaglio.17.NJP}%
  \BibitemOpen
  \bibfield  {author} {\bibinfo {author} {\bibfnamefont {M.}~\bibnamefont
  {Lostaglio}}, \bibinfo {author} {\bibfnamefont {D.}~\bibnamefont
  {Jennings}},\ and\ \bibinfo {author} {\bibfnamefont {T.}~\bibnamefont
  {Rudolph}},\ }\bibfield  {title} {\bibinfo {title} {Thermodynamic resource
  theories, non-commutativity and maximum entropy principles},\ }\href
  {https://doi.org/10.1088/1367-2630/aa617f} {\bibfield  {journal} {\bibinfo
  {journal} {New J. Phys.}\ }\textbf {\bibinfo {volume} {19}},\ \bibinfo
  {pages} {043008} (\bibinfo {year} {2017})}\BibitemShut {NoStop}%
\bibitem [{\citenamefont {Mondal}\ \emph {et~al.}()\citenamefont {Mondal},
  \citenamefont {Bhattacharyya}, \citenamefont {Ghoshal},\ and\ \citenamefont
  {Sen}}]{Mondal.23.A}%
  \BibitemOpen
  \bibfield  {author} {\bibinfo {author} {\bibfnamefont {S.}~\bibnamefont
  {Mondal}}, \bibinfo {author} {\bibfnamefont {A.}~\bibnamefont
  {Bhattacharyya}}, \bibinfo {author} {\bibfnamefont {A.}~\bibnamefont
  {Ghoshal}},\ and\ \bibinfo {author} {\bibfnamefont {U.}~\bibnamefont {Sen}},\
  }\bibfield  {title} {\bibinfo {title} {Modified {L}andauer's principle:~{H}ow
  much can the {M}axwell's demon gain by using general system-environment
  quantum state?},\ }\href {https://arxiv.org/abs/2309.09678} {\bibinfo
  {journal} {arXiv:2309.09678}\ }\BibitemShut {NoStop}%
\bibitem [{\citenamefont {van~der Meer}\ \emph {et~al.}(2022)\citenamefont
  {van~der Meer}, \citenamefont {Ertel},\ and\ \citenamefont
  {Seifert}}]{Meer.22.PRX}%
  \BibitemOpen
\bibfield  {journal} {  }\bibfield  {author} {\bibinfo {author} {\bibfnamefont
  {J.}~\bibnamefont {van~der Meer}}, \bibinfo {author} {\bibfnamefont
  {B.}~\bibnamefont {Ertel}},\ and\ \bibinfo {author} {\bibfnamefont
  {U.}~\bibnamefont {Seifert}},\ }\bibfield  {title} {\bibinfo {title}
  {Thermodynamic inference in partially accessible markov networks: A unifying
  perspective from transition-based waiting time distributions},\ }\href
  {https://doi.org/10.1103/PhysRevX.12.031025} {\bibfield  {journal} {\bibinfo
  {journal} {Phys. Rev. X}\ }\textbf {\bibinfo {volume} {12}},\ \bibinfo
  {pages} {031025} (\bibinfo {year} {2022})}\BibitemShut {NoStop}%
\bibitem [{\citenamefont {Deg\"unther}\ \emph {et~al.}(2024)\citenamefont
  {Deg\"unther}, \citenamefont {van~der Meer},\ and\ \citenamefont
  {Seifert}}]{Degunther.24.PRR}%
  \BibitemOpen
  \bibfield  {author} {\bibinfo {author} {\bibfnamefont {J.}~\bibnamefont
  {Deg\"unther}}, \bibinfo {author} {\bibfnamefont {J.}~\bibnamefont {van~der
  Meer}},\ and\ \bibinfo {author} {\bibfnamefont {U.}~\bibnamefont {Seifert}},\
  }\bibfield  {title} {\bibinfo {title} {Fluctuating entropy production on the
  coarse-grained level: Inference and localization of irreversibility},\ }\href
  {https://doi.org/10.1103/PhysRevResearch.6.023175} {\bibfield  {journal}
  {\bibinfo  {journal} {Phys. Rev. Res.}\ }\textbf {\bibinfo {volume} {6}},\
  \bibinfo {pages} {023175} (\bibinfo {year} {2024})}\BibitemShut {NoStop}%
\bibitem [{\citenamefont {Blom}\ \emph {et~al.}(2024)\citenamefont {Blom},
  \citenamefont {Song}, \citenamefont {Vouga}, \citenamefont {Godec},\ and\
  \citenamefont {Makarov}}]{Blom.24.PNAS}%
  \BibitemOpen
  \bibfield  {author} {\bibinfo {author} {\bibfnamefont {K.}~\bibnamefont
  {Blom}}, \bibinfo {author} {\bibfnamefont {K.}~\bibnamefont {Song}}, \bibinfo
  {author} {\bibfnamefont {E.}~\bibnamefont {Vouga}}, \bibinfo {author}
  {\bibfnamefont {A.}~\bibnamefont {Godec}},\ and\ \bibinfo {author}
  {\bibfnamefont {D.}~\bibnamefont {Makarov}},\ }\bibfield  {title} {\bibinfo
  {title} {Milestoning estimators of dissipation in systems observed at a
  coarse resolution},\ }\href {https://doi.org/10.1073/pnas.2318333121}
  {\bibfield  {journal} {\bibinfo  {journal} {Proc. Natl. Acad. Sci. (USA)}\
  }\textbf {\bibinfo {volume} {121}},\ \bibinfo {pages} {e2318333121} (\bibinfo
  {year} {2024})}\BibitemShut {NoStop}%
\bibitem [{Note1()}]{Note1}%
  \BibitemOpen
  \bibinfo {note} {We here explicitly refer to information-cost relations as
  those that establish fundamental bounds on thermodynamic cost in terms of the
  change in the system’s information content--quantified in the quantum
  regime by the von Neumann entropy~\cite {Lieb.02.NULL}.}\BibitemShut {Stop}%
\bibitem [{\citenamefont {Esposito}\ \emph {et~al.}(2009)\citenamefont
  {Esposito}, \citenamefont {Harbola},\ and\ \citenamefont
  {Mukamel}}]{Esposito.09.RMP}%
  \BibitemOpen
  \bibfield  {author} {\bibinfo {author} {\bibfnamefont {M.}~\bibnamefont
  {Esposito}}, \bibinfo {author} {\bibfnamefont {U.}~\bibnamefont {Harbola}},\
  and\ \bibinfo {author} {\bibfnamefont {S.}~\bibnamefont {Mukamel}},\
  }\bibfield  {title} {\bibinfo {title} {Nonequilibrium fluctuations,
  fluctuation theorems, and counting statistics in quantum systems},\ }\href
  {https://doi.org/10.1103/RevModPhys.81.1665} {\bibfield  {journal} {\bibinfo
  {journal} {Rev. Mod. Phys.}\ }\textbf {\bibinfo {volume} {81}},\ \bibinfo
  {pages} {1665} (\bibinfo {year} {2009})}\BibitemShut {NoStop}%
\bibitem [{\citenamefont {Seifert}(2012)}]{Seifert.12.RPP}%
  \BibitemOpen
  \bibfield  {author} {\bibinfo {author} {\bibfnamefont {U.}~\bibnamefont
  {Seifert}},\ }\bibfield  {title} {\bibinfo {title} {Stochastic
  thermodynamics, fluctuation theorems and molecular machines},\ }\href
  {http://stacks.iop.org/0034-4885/75/i=12/a=126001} {\bibfield  {journal}
  {\bibinfo  {journal} {Rep. Prog. Phys.}\ }\textbf {\bibinfo {volume} {75}},\
  \bibinfo {pages} {126001} (\bibinfo {year} {2012})}\BibitemShut {NoStop}%
\bibitem [{\citenamefont {Schweigler}\ \emph {et~al.}(2017)\citenamefont
  {Schweigler}, \citenamefont {Kasper}, \citenamefont {Erne}, \citenamefont
  {Mazets}, \citenamefont {Rauer}, \citenamefont {Cataldini}, \citenamefont
  {Langen}, \citenamefont {Gasenzer}, \citenamefont {Berges},\ and\
  \citenamefont {Schmiedmayer}}]{Schweigler.17.N}%
  \BibitemOpen
  \bibfield  {author} {\bibinfo {author} {\bibfnamefont {T.}~\bibnamefont
  {Schweigler}}, \bibinfo {author} {\bibfnamefont {V.}~\bibnamefont {Kasper}},
  \bibinfo {author} {\bibfnamefont {S.}~\bibnamefont {Erne}}, \bibinfo {author}
  {\bibfnamefont {I.}~\bibnamefont {Mazets}}, \bibinfo {author} {\bibfnamefont
  {B.}~\bibnamefont {Rauer}}, \bibinfo {author} {\bibfnamefont
  {F.}~\bibnamefont {Cataldini}}, \bibinfo {author} {\bibfnamefont
  {T.}~\bibnamefont {Langen}}, \bibinfo {author} {\bibfnamefont
  {T.}~\bibnamefont {Gasenzer}}, \bibinfo {author} {\bibfnamefont
  {J.}~\bibnamefont {Berges}},\ and\ \bibinfo {author} {\bibfnamefont
  {J.}~\bibnamefont {Schmiedmayer}},\ }\bibfield  {title} {\bibinfo {title}
  {Experimental characterization of a quantum many-body system via higher-order
  correlations},\ }\href {https://doi.org/10.1038/nature22310} {\bibfield
  {journal} {\bibinfo  {journal} {Nature}\ }\textbf {\bibinfo {volume} {545}},\
  \bibinfo {pages} {323} (\bibinfo {year} {2017})}\BibitemShut {NoStop}%
\bibitem [{\citenamefont {Ciliberto}(2017)}]{Ciliberto.17.PRX}%
  \BibitemOpen
  \bibfield  {author} {\bibinfo {author} {\bibfnamefont {S.}~\bibnamefont
  {Ciliberto}},\ }\bibfield  {title} {\bibinfo {title} {Experiments in
  stochastic thermodynamics: Short history and perspectives},\ }\href
  {https://doi.org/10.1103/PhysRevX.7.021051} {\bibfield  {journal} {\bibinfo
  {journal} {Phys. Rev. X}\ }\textbf {\bibinfo {volume} {7}},\ \bibinfo {pages}
  {021051} (\bibinfo {year} {2017})}\BibitemShut {NoStop}%
\bibitem [{\citenamefont {Impertro}\ \emph {et~al.}(2024)\citenamefont
  {Impertro}, \citenamefont {Karch}, \citenamefont {Wienand}, \citenamefont
  {Huh}, \citenamefont {Schweizer}, \citenamefont {Bloch},\ and\ \citenamefont
  {Aidelsburger}}]{Impertro.24.PRL}%
  \BibitemOpen
  \bibfield  {author} {\bibinfo {author} {\bibfnamefont {A.}~\bibnamefont
  {Impertro}}, \bibinfo {author} {\bibfnamefont {S.}~\bibnamefont {Karch}},
  \bibinfo {author} {\bibfnamefont {J.}~\bibnamefont {Wienand}}, \bibinfo
  {author} {\bibfnamefont {S.}~\bibnamefont {Huh}}, \bibinfo {author}
  {\bibfnamefont {C.}~\bibnamefont {Schweizer}}, \bibinfo {author}
  {\bibfnamefont {I.}~\bibnamefont {Bloch}},\ and\ \bibinfo {author}
  {\bibfnamefont {M.}~\bibnamefont {Aidelsburger}},\ }\bibfield  {title}
  {\bibinfo {title} {Local readout and control of current and kinetic energy
  operators in optical lattices},\ }\href
  {https://doi.org/10.1103/PhysRevLett.133.063401} {\bibfield  {journal}
  {\bibinfo  {journal} {Phys. Rev. Lett.}\ }\textbf {\bibinfo {volume} {133}},\
  \bibinfo {pages} {063401} (\bibinfo {year} {2024})}\BibitemShut {NoStop}%
\bibitem [{\citenamefont {Joshi}\ \emph {et~al.}(2025)\citenamefont {Joshi},
  \citenamefont {Ares}, \citenamefont {Joshi}, \citenamefont {Roos},\ and\
  \citenamefont {Calabrese}}]{Joshi.25.PRL}%
  \BibitemOpen
  \bibfield  {author} {\bibinfo {author} {\bibfnamefont {L.}~\bibnamefont
  {Joshi}}, \bibinfo {author} {\bibfnamefont {F.}~\bibnamefont {Ares}},
  \bibinfo {author} {\bibfnamefont {M.}~\bibnamefont {Joshi}}, \bibinfo
  {author} {\bibfnamefont {C.}~\bibnamefont {Roos}},\ and\ \bibinfo {author}
  {\bibfnamefont {P.}~\bibnamefont {Calabrese}},\ }\bibfield  {title} {\bibinfo
  {title} {Measuring full counting statistics in a trapped-ion quantum
  simulator},\ }\href {https://doi.org/10.1103/gyvf-s5bd} {\bibfield  {journal}
  {\bibinfo  {journal} {Phys. Rev. Lett.}\ }\textbf {\bibinfo {volume} {135}},\
  \bibinfo {pages} {160601} (\bibinfo {year} {2025})}\BibitemShut {NoStop}%
\bibitem [{\citenamefont {Jaynes}(1957)}]{Jaynes.57.PR}%
  \BibitemOpen
  \bibfield  {author} {\bibinfo {author} {\bibfnamefont {E.}~\bibnamefont
  {Jaynes}},\ }\bibfield  {title} {\bibinfo {title} {Information theory and
  statistical mechanics},\ }\href {https://doi.org/10.1103/PhysRev.106.620}
  {\bibfield  {journal} {\bibinfo  {journal} {Phys. Rev.}\ }\textbf {\bibinfo
  {volume} {106}},\ \bibinfo {pages} {620} (\bibinfo {year}
  {1957})}\BibitemShut {NoStop}%
\bibitem [{\citenamefont {Lieb}(2002)}]{Lieb.02.NULL}%
  \BibitemOpen
  \bibfield  {author} {\bibinfo {author} {\bibfnamefont {E.~H.}\ \bibnamefont
  {Lieb}},\ }\bibinfo {title} {Some convexity and subadditivity properties of
  entropy},\ in\ \href {https://doi.org/10.1007/978-3-642-55925-9_7} {\emph
  {\bibinfo {booktitle} {Inequalities: Selecta of Elliott H. Lieb}}},\ \bibinfo
  {editor} {edited by\ \bibinfo {editor} {\bibfnamefont {M.}~\bibnamefont
  {Loss}}\ and\ \bibinfo {editor} {\bibfnamefont {M.~B.}\ \bibnamefont
  {Ruskai}}}\ (\bibinfo  {publisher} {Springer Berlin Heidelberg},\ \bibinfo
  {year} {2002})\ pp.\ \bibinfo {pages} {67--79}\BibitemShut {NoStop}%
\bibitem [{\citenamefont {Majidy}\ \emph {et~al.}(2023)\citenamefont {Majidy},
  \citenamefont {Braasch}, \citenamefont {Lasek}, \citenamefont {Upadhyaya},
  \citenamefont {Kalev},\ and\ \citenamefont {Yunger~Halpern}}]{Majidy.23.NRP}%
  \BibitemOpen
  \bibfield  {author} {\bibinfo {author} {\bibfnamefont {S.}~\bibnamefont
  {Majidy}}, \bibinfo {author} {\bibfnamefont {W.}~\bibnamefont {Braasch}},
  \bibinfo {author} {\bibfnamefont {A.}~\bibnamefont {Lasek}}, \bibinfo
  {author} {\bibfnamefont {T.}~\bibnamefont {Upadhyaya}}, \bibinfo {author}
  {\bibfnamefont {A.}~\bibnamefont {Kalev}},\ and\ \bibinfo {author}
  {\bibfnamefont {N.}~\bibnamefont {Yunger~Halpern}},\ }\bibfield  {title}
  {\bibinfo {title} {Noncommuting conserved charges in quantum thermodynamics
  and beyond},\ }\href {https://doi.org/10.1038/s42254-023-00641-9} {\bibfield
  {journal} {\bibinfo  {journal} {Nat. Rev. Phys.}\ }\textbf {\bibinfo {volume}
  {5}},\ \bibinfo {pages} {689} (\bibinfo {year} {2023})}\BibitemShut {NoStop}%
\bibitem [{\citenamefont {Barato}\ and\ \citenamefont
  {Seifert}(2015)}]{Barato.15.PRL}%
  \BibitemOpen
  \bibfield  {author} {\bibinfo {author} {\bibfnamefont {A.}~\bibnamefont
  {Barato}}\ and\ \bibinfo {author} {\bibfnamefont {U.}~\bibnamefont
  {Seifert}},\ }\bibfield  {title} {\bibinfo {title} {Thermodynamic uncertainty
  relation for biomolecular processes},\ }\href
  {https://doi.org/10.1103/PhysRevLett.114.158101} {\bibfield  {journal}
  {\bibinfo  {journal} {Phys. Rev. Lett.}\ }\textbf {\bibinfo {volume} {114}},\
  \bibinfo {pages} {158101} (\bibinfo {year} {2015})}\BibitemShut {NoStop}%
\bibitem [{\citenamefont {Gingrich}\ \emph {et~al.}(2016)\citenamefont
  {Gingrich}, \citenamefont {Horowitz}, \citenamefont {Perunov},\ and\
  \citenamefont {England}}]{Gingrich.16.PRL}%
  \BibitemOpen
  \bibfield  {author} {\bibinfo {author} {\bibfnamefont {T.}~\bibnamefont
  {Gingrich}}, \bibinfo {author} {\bibfnamefont {J.}~\bibnamefont {Horowitz}},
  \bibinfo {author} {\bibfnamefont {N.}~\bibnamefont {Perunov}},\ and\ \bibinfo
  {author} {\bibfnamefont {J.}~\bibnamefont {England}},\ }\bibfield  {title}
  {\bibinfo {title} {Dissipation bounds all steady-state current
  fluctuations},\ }\href {https://doi.org/10.1103/PhysRevLett.116.120601}
  {\bibfield  {journal} {\bibinfo  {journal} {Phys. Rev. Lett.}\ }\textbf
  {\bibinfo {volume} {116}},\ \bibinfo {pages} {120601} (\bibinfo {year}
  {2016})}\BibitemShut {NoStop}%
\bibitem [{\citenamefont {Horowitz}\ and\ \citenamefont
  {Gingrich}(2019)}]{Horowitz.19.NP}%
  \BibitemOpen
  \bibfield  {author} {\bibinfo {author} {\bibfnamefont {J.}~\bibnamefont
  {Horowitz}}\ and\ \bibinfo {author} {\bibfnamefont {T.}~\bibnamefont
  {Gingrich}},\ }\bibfield  {title} {\bibinfo {title} {Thermodynamic
  uncertainty relations constrain non-equilibrium fluctuations},\ }\href
  {https://doi.org/10.1038/s41567-019-0702-6} {\bibfield  {journal} {\bibinfo
  {journal} {Nat. Phys.}\ ,\ \bibinfo {pages} {15}} (\bibinfo {year}
  {2019})}\BibitemShut {NoStop}%
\bibitem [{\citenamefont {Liu}\ and\ \citenamefont {Segal}(2019)}]{Liu.19.PRE}%
  \BibitemOpen
  \bibfield  {author} {\bibinfo {author} {\bibfnamefont {J.}~\bibnamefont
  {Liu}}\ and\ \bibinfo {author} {\bibfnamefont {D.}~\bibnamefont {Segal}},\
  }\bibfield  {title} {\bibinfo {title} {Thermodynamic uncertainty relation in
  quantum thermoelectric junctions},\ }\href
  {https://doi.org/10.1103/PhysRevE.99.062141} {\bibfield  {journal} {\bibinfo
  {journal} {Phys. Rev. E}\ }\textbf {\bibinfo {volume} {99}},\ \bibinfo
  {pages} {062141} (\bibinfo {year} {2019})}\BibitemShut {NoStop}%
\bibitem [{\citenamefont {Balian}(2007)}]{Balian.07.NULL}%
  \BibitemOpen
  \bibfield  {author} {\bibinfo {author} {\bibfnamefont {R.}~\bibnamefont
  {Balian}},\ }\href@noop {} {\emph {\bibinfo {title} {From Microphysics to
  Macrophysics: Methods and Applications of Statistical Physics}}}\ (\bibinfo
  {publisher} {Springer,Berlin Heidelberg},\ \bibinfo {year}
  {2007})\BibitemShut {NoStop}%
\bibitem [{\citenamefont {Guryanova}\ \emph {et~al.}(2016)\citenamefont
  {Guryanova}, \citenamefont {Popescu}, \citenamefont {Short}, \citenamefont
  {Silva},\ and\ \citenamefont {Skrzypczyk}}]{Guryanova.16.NC}%
  \BibitemOpen
  \bibfield  {author} {\bibinfo {author} {\bibfnamefont {Y.}~\bibnamefont
  {Guryanova}}, \bibinfo {author} {\bibfnamefont {S.}~\bibnamefont {Popescu}},
  \bibinfo {author} {\bibfnamefont {A.}~\bibnamefont {Short}}, \bibinfo
  {author} {\bibfnamefont {R.}~\bibnamefont {Silva}},\ and\ \bibinfo {author}
  {\bibfnamefont {P.}~\bibnamefont {Skrzypczyk}},\ }\bibfield  {title}
  {\bibinfo {title} {Thermodynamics of quantum systems with multiple conserved
  quantities},\ }\href {https://doi.org/10.1038/ncomms12049} {\bibfield
  {journal} {\bibinfo  {journal} {Nat. Commun.}\ }\textbf {\bibinfo {volume}
  {7}},\ \bibinfo {pages} {12049} (\bibinfo {year} {2016})}\BibitemShut
  {NoStop}%
\bibitem [{Note2()}]{Note2}%
  \BibitemOpen
  \bibinfo {note} {We emphasize that the conservation of charges in open
  systems is defined in the global sense, that is, they are preserved in the
  composite system consisting of the system of interest and its environment,
  while allowing exchange between them.}\BibitemShut {Stop}%
\bibitem [{\citenamefont {Yunger~Halpern}(2018)}]{Halpern.18.JPA}%
  \BibitemOpen
  \bibfield  {author} {\bibinfo {author} {\bibfnamefont {N.}~\bibnamefont
  {Yunger~Halpern}},\ }\bibfield  {title} {\bibinfo {title} {Beyond heat baths
  {II}: framework for generalized thermodynamic resource theories},\ }\href
  {https://doi.org/10.1088/1751-8121/aaa62f} {\bibfield  {journal} {\bibinfo
  {journal} {J. Phys. A: Math. Theor.}\ }\textbf {\bibinfo {volume} {51}},\
  \bibinfo {pages} {094001} (\bibinfo {year} {2018})}\BibitemShut {NoStop}%
\bibitem [{\citenamefont {Yunger~Halpern}\ \emph {et~al.}(2016)\citenamefont
  {Yunger~Halpern}, \citenamefont {Faist}, \citenamefont {Oppenheim},\ and\
  \citenamefont {Winter}}]{Halpern.16.NC}%
  \BibitemOpen
  \bibfield  {author} {\bibinfo {author} {\bibfnamefont {N.}~\bibnamefont
  {Yunger~Halpern}}, \bibinfo {author} {\bibfnamefont {P.}~\bibnamefont
  {Faist}}, \bibinfo {author} {\bibfnamefont {J.}~\bibnamefont {Oppenheim}},\
  and\ \bibinfo {author} {\bibfnamefont {A.}~\bibnamefont {Winter}},\
  }\bibfield  {title} {\bibinfo {title} {Microcanonical and resource-theoretic
  derivations of the thermal state of a quantum system with noncommuting
  charges},\ }\href {https://doi.org/10.1038/ncomms12051} {\bibfield  {journal}
  {\bibinfo  {journal} {Nat. Commun.}\ }\textbf {\bibinfo {volume} {7}},\
  \bibinfo {pages} {12051} (\bibinfo {year} {2016})}\BibitemShut {NoStop}%
\bibitem [{\citenamefont {Guan}\ and\ \citenamefont {Liu}(2025)}]{Guan.25.PRA}%
  \BibitemOpen
  \bibfield  {author} {\bibinfo {author} {\bibfnamefont {R.}~\bibnamefont
  {Guan}}\ and\ \bibinfo {author} {\bibfnamefont {J.}~\bibnamefont {Liu}},\
  }\bibfield  {title} {\bibinfo {title} {Anomalous flow in correlated quantum
  systems: No-go result and multiple-charge scenario},\ }\href
  {https://doi.org/10.1103/gy6n-5x26} {\bibfield  {journal} {\bibinfo
  {journal} {Phys. Rev. A}\ }\textbf {\bibinfo {volume} {112}},\ \bibinfo
  {pages} {022220} (\bibinfo {year} {2025})}\BibitemShut {NoStop}%
\bibitem [{\citenamefont {Guimar\~aes}\ \emph {et~al.}(2016)\citenamefont
  {Guimar\~aes}, \citenamefont {Landi},\ and\ \citenamefont
  {de~Oliveira}}]{Guimaraes.16.PRE}%
  \BibitemOpen
  \bibfield  {author} {\bibinfo {author} {\bibfnamefont {P.}~\bibnamefont
  {Guimar\~aes}}, \bibinfo {author} {\bibfnamefont {G.}~\bibnamefont {Landi}},\
  and\ \bibinfo {author} {\bibfnamefont {M.}~\bibnamefont {de~Oliveira}},\
  }\bibfield  {title} {\bibinfo {title} {Nonequilibrium quantum chains under
  multisite lindblad baths},\ }\href
  {https://doi.org/10.1103/PhysRevE.94.032139} {\bibfield  {journal} {\bibinfo
  {journal} {Phys. Rev. E}\ }\textbf {\bibinfo {volume} {94}},\ \bibinfo
  {pages} {032139} (\bibinfo {year} {2016})}\BibitemShut {NoStop}%
\bibitem [{\citenamefont {Lu}\ and\ \citenamefont {Sun}(2024)}]{Lu.24.PRA}%
  \BibitemOpen
  \bibfield  {author} {\bibinfo {author} {\bibfnamefont {C.}~\bibnamefont
  {Lu}}\ and\ \bibinfo {author} {\bibfnamefont {G.}~\bibnamefont {Sun}},\
  }\bibfield  {title} {\bibinfo {title} {Many-body entanglement and spectral
  clusters in the extended hard-core bosonic {H}atano-{N}elson model},\ }\href
  {https://doi.org/10.1103/PhysRevA.109.042208} {\bibfield  {journal} {\bibinfo
   {journal} {Phys. Rev. A}\ }\textbf {\bibinfo {volume} {109}},\ \bibinfo
  {pages} {042208} (\bibinfo {year} {2024})}\BibitemShut {NoStop}%
\bibitem [{\citenamefont {Herrera}\ \emph {et~al.}(2021)\citenamefont
  {Herrera}, \citenamefont {Peterson}, \citenamefont {Serra},\ and\
  \citenamefont {D'Amico}}]{Herrera.21.PRL}%
  \BibitemOpen
  \bibfield  {author} {\bibinfo {author} {\bibfnamefont {M.}~\bibnamefont
  {Herrera}}, \bibinfo {author} {\bibfnamefont {J.}~\bibnamefont {Peterson}},
  \bibinfo {author} {\bibfnamefont {R.}~\bibnamefont {Serra}},\ and\ \bibinfo
  {author} {\bibfnamefont {I.}~\bibnamefont {D'Amico}},\ }\bibfield  {title}
  {\bibinfo {title} {Easy access to energy fluctuations in nonequilibrium
  quantum many-body systems},\ }\href
  {https://doi.org/10.1103/PhysRevLett.127.030602} {\bibfield  {journal}
  {\bibinfo  {journal} {Phys. Rev. Lett.}\ }\textbf {\bibinfo {volume} {127}},\
  \bibinfo {pages} {030602} (\bibinfo {year} {2021})}\BibitemShut {NoStop}%
\bibitem [{\citenamefont {Genoni}\ \emph {et~al.}(2016)\citenamefont {Genoni},
  \citenamefont {Lami},\ and\ \citenamefont {Serafini}}]{Genoni.16.CP}%
  \BibitemOpen
  \bibfield  {author} {\bibinfo {author} {\bibfnamefont {M.}~\bibnamefont
  {Genoni}}, \bibinfo {author} {\bibfnamefont {L.}~\bibnamefont {Lami}},\ and\
  \bibinfo {author} {\bibfnamefont {A.}~\bibnamefont {Serafini}},\ }\bibfield
  {title} {\bibinfo {title} {Conditional and unconditional gaussian quantum
  dynamics},\ }\href {https://doi.org/10.1080/00107514.2015.1125624} {\bibfield
   {journal} {\bibinfo  {journal} {Contemporary Phys.}\ }\textbf {\bibinfo
  {volume} {57}},\ \bibinfo {pages} {331} (\bibinfo {year} {2016})}\BibitemShut
  {NoStop}%
\bibitem [{\citenamefont {Braunstein}\ and\ \citenamefont {van
  Loock}(2005)}]{Braunstein.05.RMP}%
  \BibitemOpen
  \bibfield  {author} {\bibinfo {author} {\bibfnamefont {S.}~\bibnamefont
  {Braunstein}}\ and\ \bibinfo {author} {\bibfnamefont {P.}~\bibnamefont {van
  Loock}},\ }\bibfield  {title} {\bibinfo {title} {Quantum information with
  continuous variables},\ }\href {https://doi.org/10.1103/RevModPhys.77.513}
  {\bibfield  {journal} {\bibinfo  {journal} {Rev. Mod. Phys.}\ }\textbf
  {\bibinfo {volume} {77}},\ \bibinfo {pages} {513} (\bibinfo {year}
  {2005})}\BibitemShut {NoStop}%
\bibitem [{\citenamefont {Weedbrook}\ \emph {et~al.}(2012)\citenamefont
  {Weedbrook}, \citenamefont {Pirandola}, \citenamefont {Garc\'{\i}a-Patr\'on},
  \citenamefont {Cerf}, \citenamefont {Ralph}, \citenamefont {Shapiro},\ and\
  \citenamefont {Lloyd}}]{Weedbrook.12.RMP}%
  \BibitemOpen
  \bibfield  {author} {\bibinfo {author} {\bibfnamefont {C.}~\bibnamefont
  {Weedbrook}}, \bibinfo {author} {\bibfnamefont {S.}~\bibnamefont
  {Pirandola}}, \bibinfo {author} {\bibfnamefont {R.}~\bibnamefont
  {Garc\'{\i}a-Patr\'on}}, \bibinfo {author} {\bibfnamefont {N.}~\bibnamefont
  {Cerf}}, \bibinfo {author} {\bibfnamefont {T.}~\bibnamefont {Ralph}},
  \bibinfo {author} {\bibfnamefont {J.}~\bibnamefont {Shapiro}},\ and\ \bibinfo
  {author} {\bibfnamefont {S.}~\bibnamefont {Lloyd}},\ }\bibfield  {title}
  {\bibinfo {title} {Gaussian quantum information},\ }\href
  {https://doi.org/10.1103/RevModPhys.84.621} {\bibfield  {journal} {\bibinfo
  {journal} {Rev. Mod. Phys.}\ }\textbf {\bibinfo {volume} {84}},\ \bibinfo
  {pages} {621} (\bibinfo {year} {2012})}\BibitemShut {NoStop}%
\bibitem [{\citenamefont {Jiang}\ and\ \citenamefont {Imry}(2016)}]{JiangCRP}%
  \BibitemOpen
  \bibfield  {author} {\bibinfo {author} {\bibfnamefont {J.-H.}\ \bibnamefont
  {Jiang}}\ and\ \bibinfo {author} {\bibfnamefont {Y.}~\bibnamefont {Imry}},\
  }\bibfield  {title} {\bibinfo {title} {Linear and nonlinear mesoscopic
  thermoelectric transport with coupling with heat baths},\ }\href
  {https://doi.org/https://doi.org/10.1016/j.crhy.2016.08.006} {\bibfield
  {journal} {\bibinfo  {journal} {C. R. Phys.}\ }\textbf {\bibinfo {volume}
  {17}},\ \bibinfo {pages} {1047 } (\bibinfo {year} {2016})}\BibitemShut
  {NoStop}%
\bibitem [{\citenamefont {Wang}\ \emph {et~al.}(2022)\citenamefont {Wang},
  \citenamefont {Wang}, \citenamefont {Lu},\ and\ \citenamefont
  {Jiang}}]{JiangReview}%
  \BibitemOpen
  \bibfield  {author} {\bibinfo {author} {\bibfnamefont {R.}~\bibnamefont
  {Wang}}, \bibinfo {author} {\bibfnamefont {C.}~\bibnamefont {Wang}}, \bibinfo
  {author} {\bibfnamefont {J.}~\bibnamefont {Lu}},\ and\ \bibinfo {author}
  {\bibfnamefont {J.-H.}\ \bibnamefont {Jiang}},\ }\bibfield  {title} {\bibinfo
  {title} {Inelastic thermoelectric transport and fluctuations in mesoscopic
  systems},\ }\href {https://doi.org/10.1080/23746149.2022.2082317} {\bibfield
  {journal} {\bibinfo  {journal} {Adv. Phys.: X}\ }\textbf {\bibinfo {volume}
  {7}},\ \bibinfo {pages} {2082317} (\bibinfo {year} {2022})}\BibitemShut
  {NoStop}%
\bibitem [{\citenamefont {Lu}\ \emph {et~al.}(2024)\citenamefont {Lu},
  \citenamefont {Wang}, \citenamefont {Ren}, \citenamefont {Wang},\ and\
  \citenamefont {Jiang}}]{LuPRB24}%
  \BibitemOpen
  \bibfield  {author} {\bibinfo {author} {\bibfnamefont {J.}~\bibnamefont
  {Lu}}, \bibinfo {author} {\bibfnamefont {Z.}~\bibnamefont {Wang}}, \bibinfo
  {author} {\bibfnamefont {J.}~\bibnamefont {Ren}}, \bibinfo {author}
  {\bibfnamefont {C.}~\bibnamefont {Wang}},\ and\ \bibinfo {author}
  {\bibfnamefont {J.-H.}\ \bibnamefont {Jiang}},\ }\bibfield  {title} {\bibinfo
  {title} {Coherence-enhanced thermodynamic performance in a periodically
  driven inelastic heat engine},\ }\href
  {https://doi.org/10.1103/PhysRevB.109.125407} {\bibfield  {journal} {\bibinfo
   {journal} {Phys. Rev. B}\ }\textbf {\bibinfo {volume} {109}},\ \bibinfo
  {pages} {125407} (\bibinfo {year} {2024})}\BibitemShut {NoStop}%
\bibitem [{\citenamefont {Jiang}\ \emph {et~al.}(2012)\citenamefont {Jiang},
  \citenamefont {Entin-Wohlman},\ and\ \citenamefont {Imry}}]{Jiang2012}%
  \BibitemOpen
  \bibfield  {author} {\bibinfo {author} {\bibfnamefont {J.-H.}\ \bibnamefont
  {Jiang}}, \bibinfo {author} {\bibfnamefont {O.}~\bibnamefont
  {Entin-Wohlman}},\ and\ \bibinfo {author} {\bibfnamefont {Y.}~\bibnamefont
  {Imry}},\ }\bibfield  {title} {\bibinfo {title} {Thermoelectric
  three-terminal hopping transport through one-dimensional nanosystems},\
  }\href {https://doi.org/10.1103/PhysRevB.85.075412} {\bibfield  {journal}
  {\bibinfo  {journal} {Phys. Rev. B}\ }\textbf {\bibinfo {volume} {85}},\
  \bibinfo {pages} {075412} (\bibinfo {year} {2012})}\BibitemShut {NoStop}%
\bibitem [{\citenamefont {Lu}\ \emph {et~al.}(2023)\citenamefont {Lu},
  \citenamefont {Wang}, \citenamefont {Wang}, \citenamefont {Peng},
  \citenamefont {Wang},\ and\ \citenamefont {Jiang}}]{MyPRBMultitask}%
  \BibitemOpen
  \bibfield  {author} {\bibinfo {author} {\bibfnamefont {J.}~\bibnamefont
  {Lu}}, \bibinfo {author} {\bibfnamefont {Z.}~\bibnamefont {Wang}}, \bibinfo
  {author} {\bibfnamefont {R.}~\bibnamefont {Wang}}, \bibinfo {author}
  {\bibfnamefont {J.}~\bibnamefont {Peng}}, \bibinfo {author} {\bibfnamefont
  {C.}~\bibnamefont {Wang}},\ and\ \bibinfo {author} {\bibfnamefont {J.-H.}\
  \bibnamefont {Jiang}},\ }\bibfield  {title} {\bibinfo {title} {Multitask
  quantum thermal machines and cooperative effects},\ }\href
  {https://doi.org/10.1103/PhysRevB.107.075428} {\bibfield  {journal} {\bibinfo
   {journal} {Phys. Rev. B}\ }\textbf {\bibinfo {volume} {107}},\ \bibinfo
  {pages} {075428} (\bibinfo {year} {2023})}\BibitemShut {NoStop}%
\bibitem [{\citenamefont {Breuer}\ and\ \citenamefont
  {Petruccione}(2002)}]{Breuer.02.NULL}%
  \BibitemOpen
  \bibfield  {author} {\bibinfo {author} {\bibfnamefont {H.-P.}\ \bibnamefont
  {Breuer}}\ and\ \bibinfo {author} {\bibfnamefont {F.}~\bibnamefont
  {Petruccione}},\ }\href@noop {} {\emph {\bibinfo {title} {The Theory of Open
  Quantum Systems}}}\ (\bibinfo  {publisher} {Oxford University Press, New
  York},\ \bibinfo {year} {2002})\BibitemShut {NoStop}%
\bibitem [{\citenamefont {Harrington}\ \emph {et~al.}(2022)\citenamefont
  {Harrington}, \citenamefont {Mueller},\ and\ \citenamefont
  {Murch}}]{Harrington.22.NRP}%
  \BibitemOpen
  \bibfield  {author} {\bibinfo {author} {\bibfnamefont {P.}~\bibnamefont
  {Harrington}}, \bibinfo {author} {\bibfnamefont {E.}~\bibnamefont
  {Mueller}},\ and\ \bibinfo {author} {\bibfnamefont {K.}~\bibnamefont
  {Murch}},\ }\bibfield  {title} {\bibinfo {title} {Engineered dissipation for
  quantum information science},\ }\href
  {https://doi.org/10.1038/s42254-022-00494-8} {\bibfield  {journal} {\bibinfo
  {journal} {Nat. Rev. Phys.}\ ,\ \bibinfo {pages} {660}} (\bibinfo {year}
  {2022})}\BibitemShut {NoStop}%
\end{thebibliography}

\end{document}